\documentclass[twocolumn]{aastex631}

\usepackage[utf8]{inputenc}
\usepackage{amssymb}
\usepackage{newunicodechar}

\begin{document}

\title{The JCMT BISTRO Survey: A Spiral Magnetic Field in a Hub-Filament Structure, Monoceros R2}

\author[0000-0001-7866-2686]{Jihye Hwang}
\email{hjh3772@gmail.com}
\affil{Korea Astronomy and Space Science Institute (KASI), 776 Daedeokdae-ro, Yuseong-gu, Daejeon 34055, Republic of Korea}
\affil{University of Science and Technology, Korea (UST), 217 Gajeong-ro, Yuseong-gu, Daejeon 34113, Republic of Korea}

\author[0000-0002-1229-0426]{Jongsoo Kim}
\affil{Korea Astronomy and Space Science Institute (KASI), 776 Daedeokdae-ro, Yuseong-gu, Daejeon 34055, Republic of Korea}
\affil{University of Science and Technology, Korea (UST), 217 Gajeong-ro, Yuseong-gu, Daejeon 34113, Republic of Korea}

\author[0000-0002-8557-3582]{Kate Pattle}
\affil{Department of Physics and Astronomy, University College London, WC1E 6BT London, UK}

\author[0000-0002-3179-6334]{Chang Won Lee}
\affil{Korea Astronomy and Space Science Institute (KASI), 776 Daedeokdae-ro, Yuseong-gu, Daejeon 34055, Republic of Korea}
\affil{University of Science and Technology, Korea (UST), 217 Gajeong-ro, Yuseong-gu, Daejeon 34113, Republic of Korea}

\author[0000-0003-2777-5861]{Patrick M. Koch}
\affil{Academia Sinica Institute of Astronomy and Astrophysics, No.1, Sec. 4., Roosevelt Road, Taipei 10617, Taiwan}

\author[0000-0002-6773-459X]{Doug Johnstone}
\affil{NRC Herzberg Astronomy and Astrophysics, 5071 West Saanich Road, Victoria, BC V9E 2E7, Canada}
\affil{Department of Physics and Astronomy, University of Victoria, Victoria, BC V8W 2Y2, Canada}

\author{Kohji Tomisaka}
\affil{Division of Theoretical Astronomy, National Astronomical Observatory of Japan, Mitaka, Tokyo 181-8588, Japan}
\affil{SOKENDAI (The Graduate University for Advanced Studies), Hayama, Kanagawa 240-0193, Japan}

\author[0000-0002-1178-5486]{Anthony Whitworth}
\affil{School of Physics and Astronomy, Cardiff University, The Parade, Cardiff, CF24 3AA, UK}

\author[0000-0003-0646-8782]{Ray S. Furuya}
\affil{Institute of Liberal Arts and Sciences Tokushima University, Minami Jousanajima-machi 1-1, Tokushima 770-8502, Japan}

\author[0000-0001-7379-6263]{Ji-hyun Kang}
\affil{Korea Astronomy and Space Science Institute (KASI), 776 Daedeokdae-ro, Yuseong-gu, Daejeon 34055, Republic of Korea}

% ...

\begin{abstract}

We present and analyze observations of  polarized dust emission at 850 $\mu$m  towards the central 1 pc $\times$ 1 pc hub-filament structure of Monoceros R2 (Mon R2). The data are obtained with SCUBA-2/POL-2 on the James Clerk Maxwell Telescope (JCMT) as part of the BISTRO (B-fields in Star-forming Region Observations) survey. The orientations of the magnetic field follow the spiral structure of Mon R2, which are well-described by an axisymmetric magnetic field model. We estimate the turbulent component of the magnetic field using the angle difference between our observations and the best-fit model of the underlying large-scale mean magnetic field. This estimate is used to calculate the magnetic field strength using the Davis-Chandrasekhar-Fermi method, for which we also obtain the distribution of volume density and velocity dispersion using a column density map derived from $Herschel$ data and the C$^{18}$O ($J$ = 3-2) data taken with HARP on the JCMT, respectively. We make maps of magnetic field strengths and mass-to-flux ratios, finding that magnetic field strengths vary from 0.02 to 3.64 mG with a mean value of 1.0 $\pm$ 0.06 mG, and the mean critical mass-to-flux ratio is 0.47 $\pm$ 0.02. Additionally, the mean Alfv\'en Mach number is 0.35 $\pm$ 0.01. This suggests that in Mon R2, magnetic fields provide resistance against large-scale gravitational collapse, and magnetic pressure exceeds turbulent pressure. We also investigate the properties of each filament in Mon R2. Most of the filaments are aligned along the magnetic field direction and are magnetically sub-critical.

\end{abstract}
\keywords{Star Formation (1569) --- Interstellar Medium (847) --- Magnetic fields (994)}


\section{introduction} \label{p2:sec:intro}
Hub-filament Systems (HFSs) are ubiquitous in star-forming regions and play an important role in the formation and evolution of high-mass stars and star clusters (\citealt{Myers2009}; \citealt{Peretto2013}; \citealt{Kumar2020}). In HFSs, filaments are elongated structures with high aspect ratios, while hubs, which are located at the junctions of filaments, have larger column densities and low aspect ratios (\citealt{Myers2009}; \citealt{Pretto2014}). The longitudinal mass flow along filaments increases the density of hubs and hence increases the level of star-formation activity in the hubs (\citealt{Pretto2014}; \citealt{Trevino2019}; \citealt{Pillai2020}).
Recently, \citet{Kumar2020} found 3700 HFS candidates using the \textit{Herschel} Hi-GAL Survey catalogue, and suggested a four-stage Filaments to Clusters (F2C) paradigm for star-formation in a HFS. In the first stage, individual filaments approach each other. Then, the filaments overlap with each other, making a rotating flattened hub. In the third stage, young OB stars and massive cores are formed in the hub. Finally, the expanding radiation bubbles driven by OB stars create H{\sc ii} regions and 
burn out the tips of the filaments connected to the bubbles.
In this scenario, first the low mass stars form inside individual filaments and massive stars form later inside the hub, hence star clusters form in HFSs. \citet{Kumar2020} speculated that the flow of matter and the increase in the density at the hub can result in increased magnetic field strength at the hub providing support against gravity. But observational evidences of magnetic fields in HFSs are few and rare, and almost none in spatially well resolved  hubs of HFSs.

Understanding the role of magnetic fields is crucial to understand the nature of star formation in HFSs.
The physical and chemical properties and energetics of HFSs have been studied using dust continuum and spectral line data (e.g., \citealt{Trevino2019}; \citealt{Chung2021}), but there are only a few studies which have investigated the importance of magnetic fields using observations of dust polarization at far-infrared and submillimeter/millimeter (sub-mm/mm) wavelengths (e.g., \citealt{Wang2019}; \citealt{Pillai2020}; \citealt{Wang2020}; \citealt{Beuther2020} \citealt{Arzoumanian2021}).
\citet{Pillai2020} suggested that magnetized gas flows along filaments can feed the hub in the Serpens South region. They showed that magnetic field orientations in this region were parallel to the filaments and perpendicular to the hub. \citet{Beuther2020} showed that magnetic field structures can be also shaped by gravitational contraction and rotation in the high-mass star-forming region G327.3. Polarization observations in another high-mass region, G31.41+0.31, revealed an hourglass  magnetic field morphology \citep{Girart2009}, while high-resolution ALMA observations showed  the detailed magnetic field morphology, finding radially-converging field lines \citep{Beltran2019}. \citet{Beltran2019} interpreted this magnetic field morphology using a model in which an axially-symmetric singular toroid is threaded by a poloidal magnetic field.
However, these studies of magnetic field structures are insufficient to draw a conclusive picture on the importance of magnetic fields in HFSs.

It is necessary to measure magnetic field strengths within HFSs in order to quantify their role. The Davis-Chandrasekhar-Fermi (DCF; \citealt{Davis1951}; \citealt{ChanFer1953}) method has generally been used to estimate magnetic field strengths using polarized dust emission in star-forming regions, including in HFSs. \citet{Wang2020} estimated magnetic field strengths in the HFS, G33.92+0.11, using the DCF method. They compared relative importance of gravitational, magnetic, and kinematic energies in the HFS, finding that the magnetic and gravitational energies are the largest and the smallest terms in the energy balance, respectively.
The DCF method has been applied to a few HFSs, and their mean magnetic field strengths have been obtained (e.g., \citealt{Wang2020}). However, the mean field strength within a HFS may not provide enough information to determine the importance of the magnetic field across the HFS. HFSs have a complex structure containing turbulent motion, radiative feedback by formation of OB stars, infall motion or rotation, which may affect magnetic field structures within the HFS. \citet{Hwang2021} and \citet{Guerra2021} suggested new ways to estimate the spatial distribution of magnetic field strengths using extensions of the unsharp-masking \citep{Pattle2017} and the structure function (\citealt{Hildebrand2009}; \citealt{Houde2009}) approaches to the DCF method, respectively. We apply the method suggested by \citet{Hwang2021} to a HFS in order to obtain the distribution of magnetic field strengths across the region allowing us to better quantify the role of magnetic fields in the HFS.

Monoceros R2 (Mon R2) is a good target to study the role of magnetic fields in HFSs, because it is one of the nearest HFSs, and therefore spatially well resolved.
Mon R2 contains features of the second or third stage of the F2C paradigm suggested by \citet{Kumar2020}. There are several filaments converging near the IRS 1 source in the central hub of Mon R2 (\citealt{Rayner2017}; \citealt{Trevino2019}). 
Mon R2 is the closest ultracompact (UC) H{\sc ii} region, at a distance of 830 pc \citep{Herbst1976}. Its physical and chemical properties have been studied using molecular lines and continuum emission with relatively high resolutions (e.g., \citealt{Didelon2015}; \citealt{Rayner2017}; \citealt{Trevino2019}; \citealt{Kumar2021}). 
\citet{Trevino2019} suggested the hub in Mon R2 may have a rotating flattened structure, based on analysis of velocity gradients and a position-velocity diagram of C$^{18}$O and $^{13}$CO (2-1) data. From the velocity gradient, they also inferred longitudinal gas flows along the filaments transferring matter toward the hub of Mon R2. They hypothesized that because of the conservation of angular momentum in a rotating cloud, large-scale radial infall of gas has been converted into a rotating flattened structure around the hub. In the hub, velocities vary with the distance to the center following a power-law relation. They showed spiral and ring structures around the IRS 1 in integrated C$^{18}$O and $^{13}$CO intensity maps, respectively. The position-velocity diagram cutting along the ring pattern showed a sinusoidal pattern, which could result from rotation.
Based on these features, they estimated the large-scale rotation of the hub in Mon R2.
The number of point sources extracted by spectral energy distribution (SED) of \textit{Herschel} data
is larger in the hub than in the filaments \citep{Rayner2017}, and so star formation is likely more active in the hub than in the filaments.
\citet{Kumar2021} decomposed diffuse cloud material, filaments and compact sources from the column density map obtained from \textit{Herschel} observations of Mon R2 using the $getsf$  algorithm \citep{Menshchikov2021}. 
They estimated a star formation efficiency of a few per cent, which is lower than that expected from predictions.  
This is possibly because the magnetic fields in Mon R2 might delay star formation, even though there is gas flow from the filaments to the hub. 
In this paper we address, by measuring the distribution of the magnetic fields from dust polarization observations, whether or not magnetic fields delay star formation or support Mon R2 against gravitational collapse.
 
Mon R2 has been observed as part of the second B-fields In STar-forming Region Observations (BISTRO) survey \citep{WardThompson2017} using the POL-2 polarimeter \citep{Friberg2016} on the Submillimetre Common-User Bolometer Array 2 (SCUBA-2; \citealt{Holland2013}) camera on the James Clerk Maxwell Telescope (JCMT). POL-2 observations have provided dust polarization maps of molecular clouds with the highest resolution achievable with currently operational single-dish radio telescopes at sub-mm wavelengths. The  initial aim of the BISTRO survey was to study magnetic fields in nearby star-forming clouds and cores. The BISTRO-2 and BISTRO-3 survey extensions aim to make observations of magnetic fields in high-mass star-forming regions and to investigate various evolutionary stages of star formation. In total, the survey is targeting 48 observing fields. The survey results have shown magnetic field structures and their properties in Orion A (\citealt{WardThompson2017}; \citealt{Pattle2017}; \citealt{Hwang2021}), Ophiuchus (\citealt{Kwon2018}; \citealt{Soam2018}; \citealt{Liu2019}; \citealt{Pattle2021}), M16 \citep{Pattle2018} IC 5146 \citep{Wang2019}, Barnard 1 \citep{Coude2019}, NGC 1333 \citep{Doi2020}, NGC 6334 \citep{Arzoumanian2021}, Rosette \citep{Konyves2021}, Auriga-California \citep{Ngoc2021}, Taurus \citep{Eswaraiah2021}, Orion B \citep{Lyo2021} and Serpens \citep{Kwon2022}. BISTRO survey data have also been used to study the polarization properties of dust grains \citep{Pattle2019}, multi-wavelength polarized emission in Orion B \citep{Fanciullo2022} and the alignment between magnetic fields and outflows in cores (\citealt{Yen2021}, \citealt{Gupta2022}).

Here, we study the role of magnetic fields in Mon R2 using polarized dust emission and C$^{18}$O molecular line observations. Although Mon R2 provides a good environment for the formation of massive stars and star clusters, the role of magnetic fields in their formation has not yet been studied. In this study, we show the magnetic field structure obtained by POL-2 on the JCMT at 850 $\mu$m. A mean magnetic field structure in Mon R2 is estimated using a rotating axisymmetric magnetic field model \citep{Wardle1990}. Using this model, we derive the distribution of magnetic field strengths using the method suggested by \citet{Hwang2021}. Additionally, we obtain maps of mass-to-flux ratios and Alfv\'en Mach number in order to compare the relative importance of magnetic fields, gravity and turbulence. We also estimate their physical properties of, as well as the magnetic field strengths along, the filaments which are identified by \citet{Kumar2021} in the \textit{Herschel} column density map of Mon R2. Based on these results, we discuss the role of magnetic fields in Mon R2. 

This paper is organized as follows: in section \ref{p2:sec:obs}, we describe the observations of Mon R2 using SCUBA-2/POL-2 and the Heterodyne Array Receiver Program (HARP) on the JCMT, and their data reduction. In section \ref{p2:sec:res}, we show the magnetic field orientations inferred from these observations. We also describe the identified filaments and their physical properties.
The magnetic properties of Mon R2: magnetic field strength, mass-to-flux ratio and Alfv\'en Mach number, are discussed in section \ref{p2:sec:dis}. In the same section, we also discuss the magnetic and other physical properties of the identified filaments. Our conclusions are given in section \ref{p2:sec:conc}.

\section{Observations} \label{p2:sec:obs}

We observed polarized dust continuum emission and C$^{18}$O J=(3-2) spectral line emission in Mon R2 using SCUBA-2/POL-2 and HARP on the JCMT, respectively. Here, we describe the two set of observations and their data reduction procedures.

\subsection{SCUBA-2/POL-2 observations} \label{p2:subsec:pol2}

 Mon R2 was observed with SCUBA-2/POL-2 as part of the BISTRO-2 large program (project code: M17BL011). Mon R2 was observed 20 times between 2017 November 24 and 2019 October 5. The observation time of each data set is $\sim$ 40 minutes. The total on-source time of the observations is about 14 hours. The observations were performed using the POL-2 DAISY observing mode \citep{Friberg2016}. The beam size at 850 $\mu$m is 14.1$''$. The observed data sets were obtained in Band 1 weather condition in which the atmospheric opacity at 225 GHz, $\tau_{225 \mathrm{GHz}}$, is less than 0.05. 
 
 The data reduction of the 20 data sets at 450 and 850 $\mu$m was conducted using the $pol2map$ routine\footnote{\url{http://starlink.eao.hawaii.edu/docs/sun258.htx/sun258ss75.html}} in the Sub-Millimetre User Reduction Facility (SMURF) Starlink software package \citep{Jenness2013} and the August 2019\footnote{\url{https://www.eaobservatory.org/jcmt/2019/08/new-ip-models-for-pol2-data/}} instrumental polarization model. We followed the data reduction process described  by \citet{Hwang2021}. The final Stokes $I$, $Q$, and $U$ maps are gridded to 4$''$ pixels and have units of pW. We converted the data units from pW to Jy beam$^{-1}$ by multiplying them by the flux conversion factor of SCUBA-2 at 850$\mu$m, 495 Jy beam$^{-1}$ pW$^{-1}$, which has recently been updated by \citet{Mairs2021}, multiplied by the usual transmission factor of 1.35 for POL-2 at 850$\mu$m \citep{Friberg2016}. The root-mean-square (rms) noise values in the Stokes $I$, $Q$, and $U$ maps are 3.4, 2.9, and 2.9 mJy beam$^{-1}$, respectively. As part of the data reduction procedure, a polarization vector catalogue is created from these Stokes maps. The polarization angle ($\theta_{obs}$), debiased polarization intensities ($PI$) and debiased polarization fraction ($p$) are obtained using the Stokes parameters, $\theta_{obs} =1/2 \arctan(U/Q)$, $PI=\sqrt{Q^2+U^2-\sigma_{PI}^2}$, and $p = (Q^2 + U^2 - 0.5[(\delta Q)^2+(\delta U)^2])^{1/2}/I$, where $\sigma_{PI}$ is the uncertainties of $PI$, $\sigma_{PI}=\sqrt{((Q\delta Q)^2+(U\delta U)^2}/(Q^2+U^2)$, and $(\delta Q)^2$ and $(\delta U)^2$ are the variances of Stokes $Q$ and $U$, respectively. The uncertainties on $\theta_{obs}$ and $p$ are calculated as $\sigma_{\theta_{obs}}=0.5\sqrt{(U\delta Q)^2+(Q\delta U)^2}/(Q^2+U^2)$ and $\sigma_p$=$\sqrt{\sigma_{PI}^2/I^2 + (\delta I^2(Q^2+U^2))/I^4}$, where $\delta I^2$ is the variance of Stokes $I$.

\begin{figure*}[htb!]
\epsscale{1.1}
\plotone{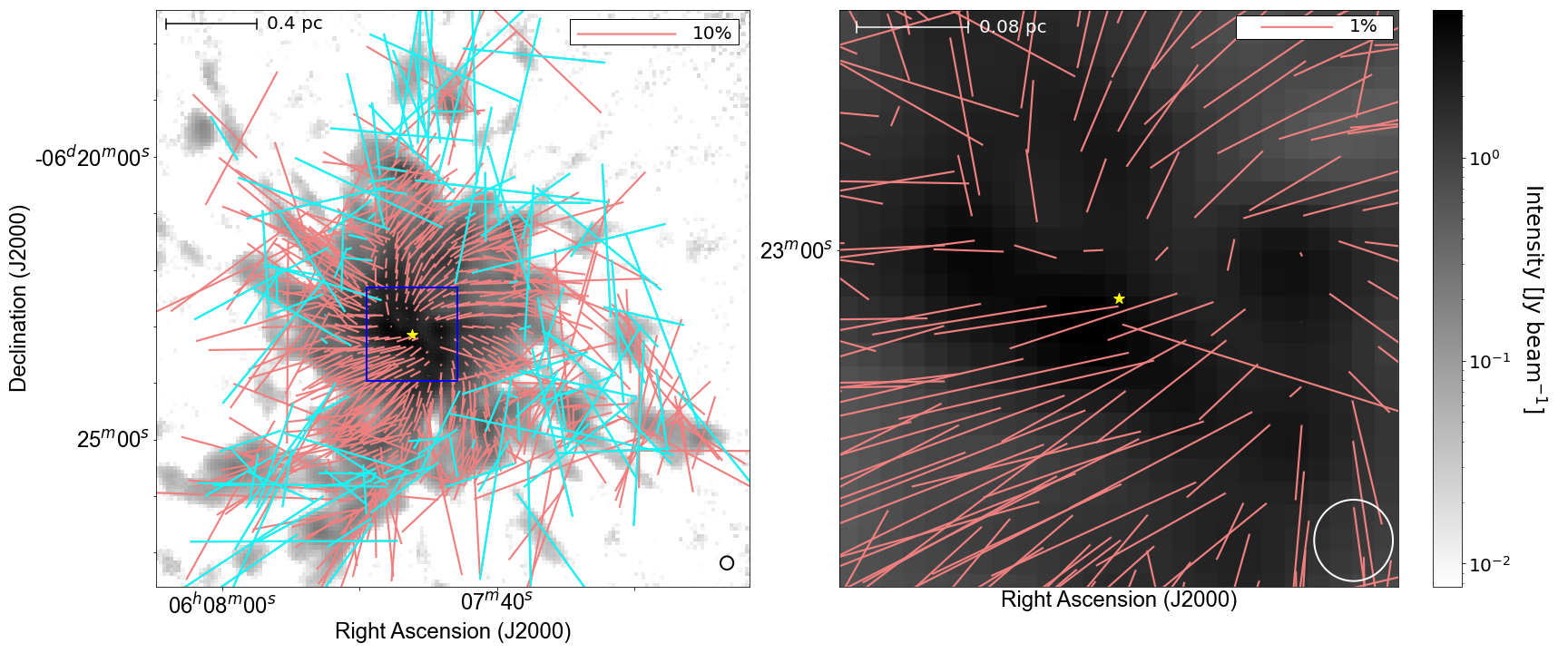}
\caption{Maps of polarization segments rotated by 90 degrees to show magnetic field orientation in Mon R2. The right panel shows a zoomed-in map of the region marked with a blue box in the left panel. The background gray-scale image shows dust continuum emission at 850 $\mu$m. The intensity scale of the image is shown as a color bar. The polarization segment selection criteria are $I$/$\delta I \geq$ 10, $p$/$\delta p \geq$ 2, $p$ $<$ 20\% and $\delta \theta_{obs}$ $<$ 15 degrees (definitions of $I$, $p$, $\delta I$, $\delta p$ and $\delta \theta_{obs}$ are given in the text). The coral and cyan segments correspond to measurements for which $p$/$\delta p \geq$ 3 and 3 $>$ $p$/$\delta p \geq$ 2, respectively. The segments shown are binned to a 12$''$ pixel grid. The lengths of the segments are scaled to $p$ and scale bars with $p=$ 10\% and 1\% are shown in the upper right corner of each panel. Yellow stars indicate the position of IRS 1 (R.A.(J2000) = 06$^{\text{h}}$07$^{\text{m}}$46.2$^{\text{s}}$, Dec.(J2000) = -06$^{\circ}$23$'$08.3$''$). The circles in the lower right corner of each panel show the JCMT beam size of 14.1$''$ at 850 $\mu$m. Physical scale bars at the distance to Mon R2 of 830 pc are shown in the upper left corner of both panels.
\label{p2:fig:pol1}}
\end{figure*}

 Figure \ref{p2:fig:pol1} shows the polarization maps of Mon R2 which we obtained. The polarization segments in the figure are rotated by 90 degrees to show the magnetic field orientations in Mon R2. The selection criteria for the polarization segments shown are $I$/$\delta I \geq$ 10, $p$/$\delta p \geq$ 2, $p$ $<$ 20\% and $\delta \theta_{obs}$ $<$ 15 degrees, where $I$ and $p$ are total intensity and polarization fraction, and $\delta I$, $\delta p$ and $\delta \theta_{obs}$ are the uncertainties on $I$, $p$ and polarization angle $\theta_{obs}$, respectively. We chose the polarization fraction selection criterion to be $p$/$\delta p \geq$ 2 in order to include segments in the outer parts of the filaments. The other criteria are widely used in previous studies using POL-2. The polarization segments within the blue box have lower polarization fractions than those elsewhere. In order to clearly display these polarization segments, a zoomed-in map of the region within the blue box is shown in the right panel of figure \ref{p2:fig:pol1}. The lengths of the polarization segments shown in both panels are scaled by $p$, scale bars of which are shown in the upper right corners of each panel. We binned the polarization segments to a 12$''$ pixel grid, which is close to the JCMT beam size at 850 $\mu$m.
 
\begin{figure*}[htb!]
\epsscale{0.7}
\plotone{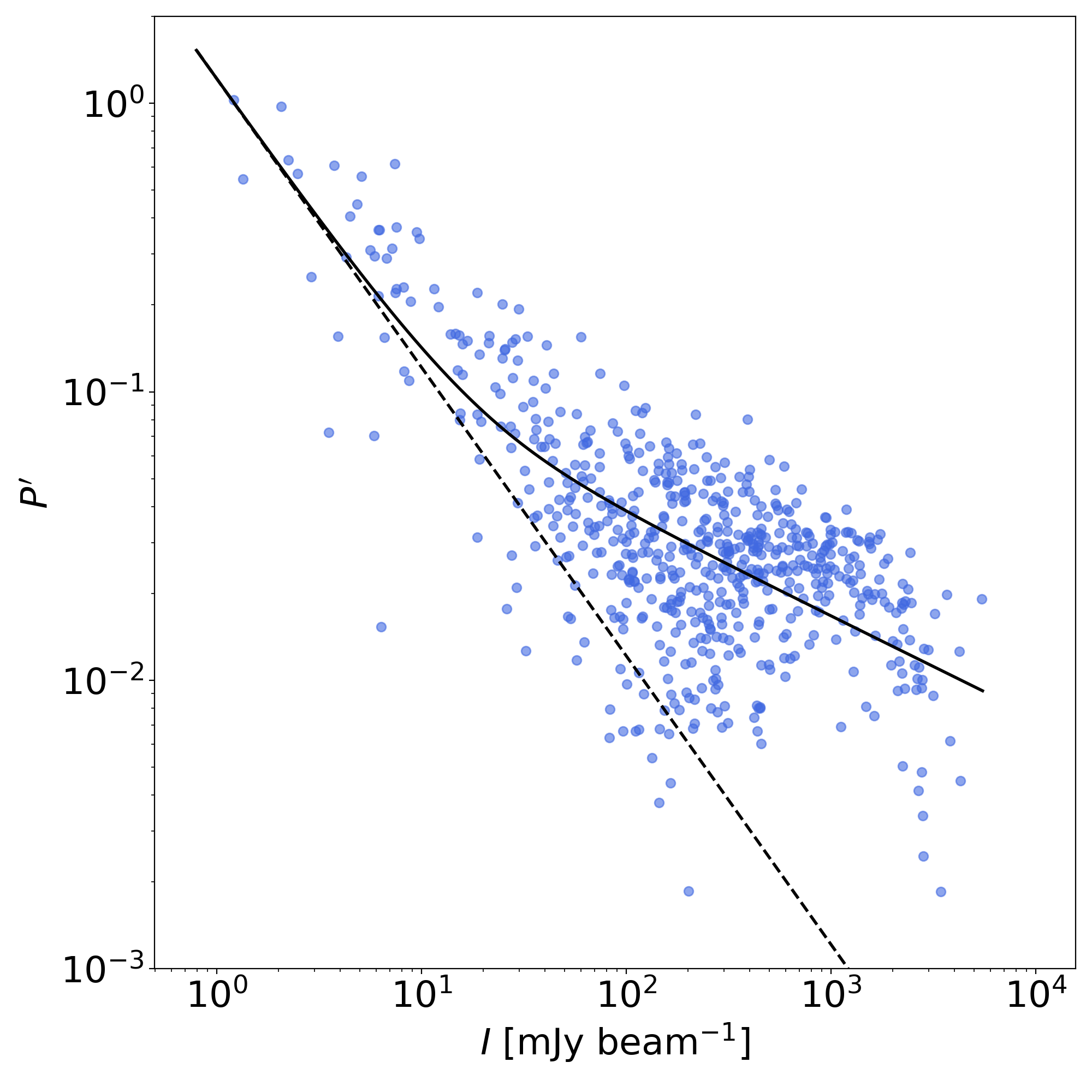}
\caption{Non-debiased polarization fraction as a function of intensity. The solid line shows the best-fit power-law model with an index $\alpha=0.35$ and Ricean noise. The dashed black line has a power-law index of $\alpha=1$, representing pure Ricean noise.
\label{p2:fig:pvsi}} 
\end{figure*}

Figure \ref{p2:fig:pvsi} shows the non-debiased polarization fraction ($p^\prime$) as a function of intensity ($I$). We plotted all polarization segments within a circle with a radius of 3$'$ from the center of the intensity map, because the POL-2 observations show the lowest, and the most consistent, noise level in this area. The solid black line is fitted using the Ricean distribution model described by \citet{Pattle2019}. The relation between polarization fraction and intensity has been modelled using a power-law, $p \propto I ^{-\alpha}$, and a Ricean noise distribution. The index $\alpha$ can be used to infer grain alignment efficiency. An index $\alpha$ $\sim$ 1 indicates a lack of alignment between dust grains and magnetic fields, and is shown as a dashed line in the figure. However, the non-Gaussian noise properties of polarization fraction, a defined-positive quantity, can cause the index of the power-law to be overestimated if a power-law model is directly fitted to the data. \citet{Pattle2019} suggested the Ricean distribution model as a means of fitting the relation between $p$ and $I$. Using this model we obtained an index of $\alpha = 0.35$, which indicates that dust grains are aligned with respect to magnetic fields in Mon R2.

\subsection{HARP observations}
\begin{figure*}[htb!]
\epsscale{0.8}
\plotone{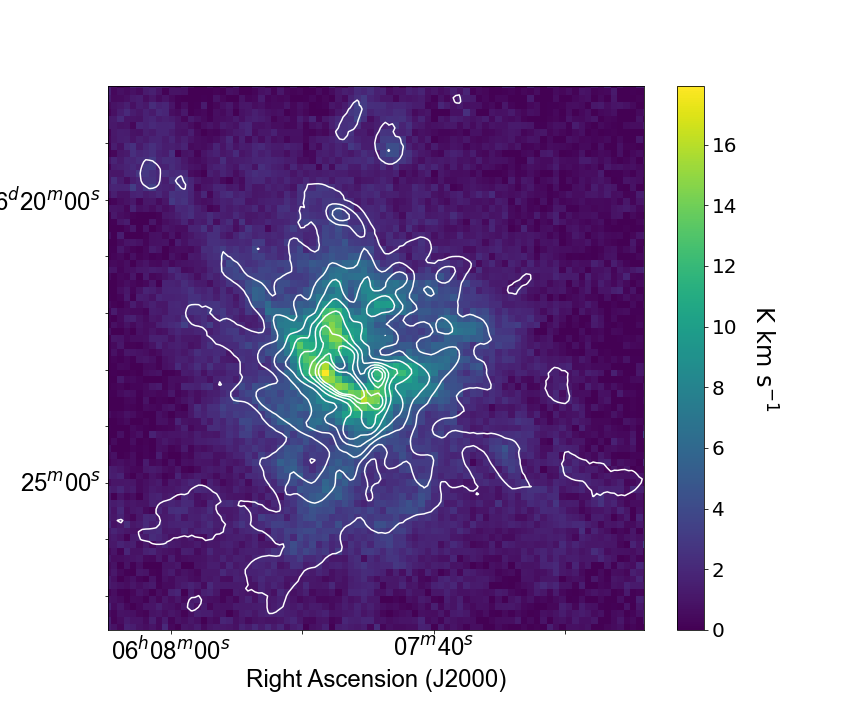}
\caption{JCMT HARP C$^{18}$O ($J$ = 3-2) integrated intensity map in Mon R2. The emission is integrated over the local standard of rest (LSR) velocity from 6 km s$^{-1}$ to 14 km s$^{-1}$. The contours show SCUBA-2 850 $\mu$m flux densities of 0.06, 0.3, 0.6, 1.0, 1.5, 2, 2.5 and 3 Jy beam $^{-1}$. \label{p2:fig:mom0}} 
\end{figure*}

C$^{18}$O ($J$ = 3-2) observations of Mon R2 were made with HARP on the JCMT under project code R19BP001 (PI: Jihye Hwang). The C$^{18}$O spectral line observations, with a 329.331 GHz, were taken between October and December 2019. The observations were performed in Band 3 weather conditions. The mean system temperatures of the data vary from 452 to 1549 K.
The C$^{18}$O data were reduced using the ORAC Data Reduction (ORAC-DR) pipeline and the Kernel Application Package (KAPPA; \citealt{Currie2008}), both of which are part of the Starlink software suite \citep{Jenness2013}. The pixel size and spectral resolution of the reduced map are 7$''$ and 0.05 km s$^{-1}$, respectively. To increase the signal-to-noise ratio (SNR), we smoothed the data cube to a spectral resolution of 0.15 km s$^{-1}$. Figure \ref{p2:fig:mom0} shows the C$^{18}$O integrated intensity in Mon R2, which is consistent with the dust continuum emission at 850 $\mu$m. 
 
\section{Results}\label{p2:sec:res}

\begin{figure*}[htb!]
\epsscale{1.1}
\plotone{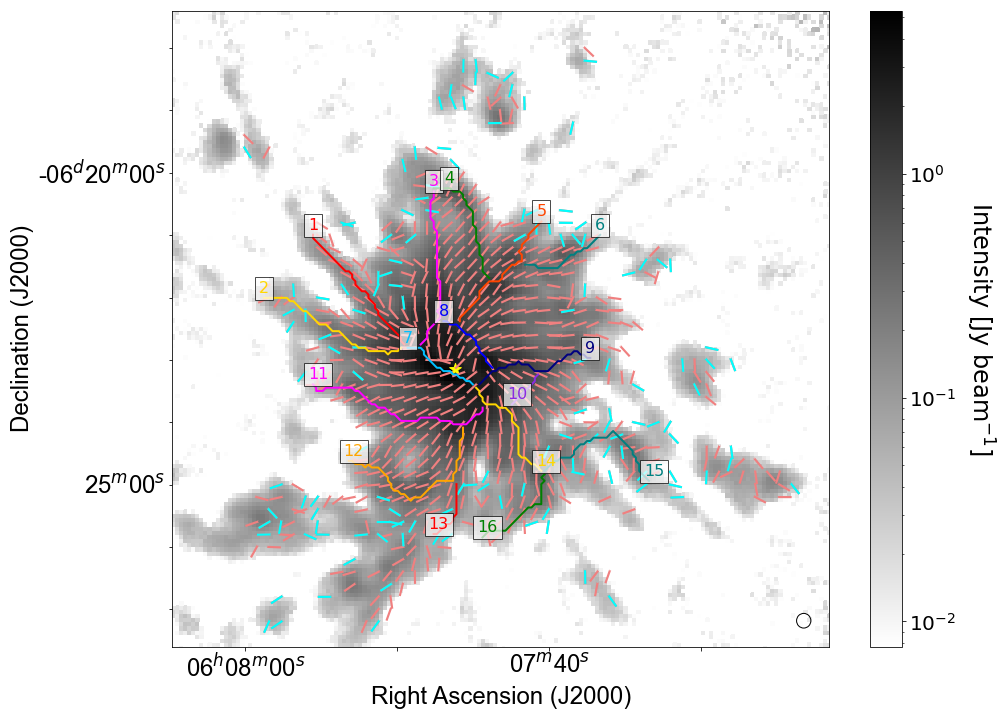}
\caption{Map of magnetic field orientation in Mon R2. 
Segments are the same in the left panel of Figure \ref{p2:fig:pol1}, but are shown with a uniform length for clarity.
 The background gray-scale image, the yellow star and the circle marking the beam are as those shown in Figure \ref{p2:fig:pol1}. Colored lines show the skeletons of the filaments which were obtained using the $getsf$ algorithm \citep{Menshchikov2021} by \citet{Kumar2021}. The numbers of each filament are marked on each rectangular box at the edges of each skeleton.
\label{p2:fig:pol}} 
\end{figure*}

\begin{figure*}[htb!]
\epsscale{1.1}
\plotone{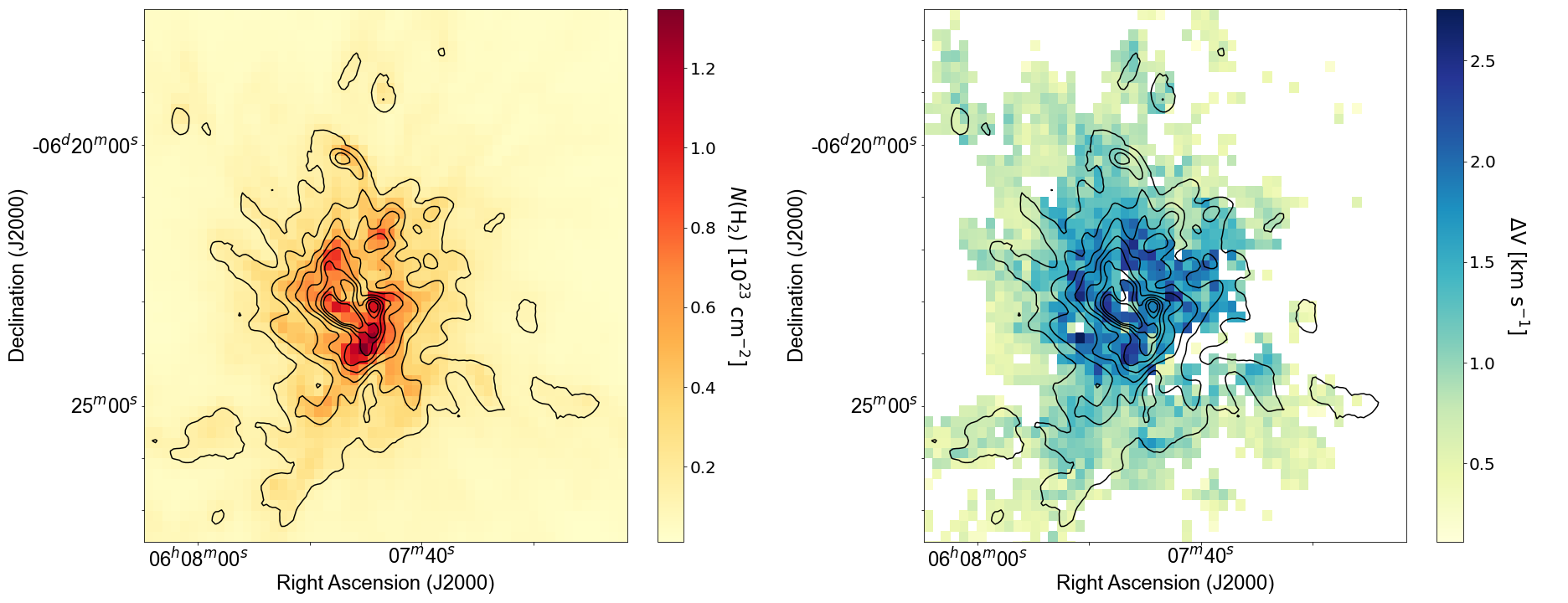}
\caption{Maps of column density (left panel) and velocity dispersion (right panel). The 18.2$''$ resolution column density map is produced by \citep{Kumar2021} using $getsf$ method. The velocity dispersion map is obtained by measuring the FWHMs of the non-thermal component of the C$^{18}$O spectral line, observed using HARP on the JCMT. Contours are the same as those defined in Figure \ref{p2:fig:mom0}. \label{p2:fig:vol_vel}} 
\end{figure*}

Figure \ref{p2:fig:pol} shows the orientation of the magnetic field within Mon R2, with segments scaled to a uniform length. The overall magnetic field appears to have a spiral structure around the IRS 1 source, which is marked with a yellow star in the figure. The magnetic field structure is discussed further in Section~\ref{p2:subsec:morphology}. The background image of Figure~\ref{p2:fig:pol} is the dust continuum emission at 850 $\mu$m (Stokes $I$). The figure also shows the 16 skeletons of filaments obtained by \citet{Kumar2021} from a high-resolution (18.2$''$) column density map of the 
component of filaments (Figure \ref{p2:fig:vol_vel}) in Mon R2 derived by the source and filament extraction method $getsf$ \citep{Menshchikov2021}. The method separates structural components of sources, filaments, and backgrounds, which allows a proper analysis. In our paper, we used the separated image of filaments that has no contribution of sources or backgrounds. The high-resolution column density image was computed from the \textit{Herschel} images of Mon R2 at 160, 250, 350, and 500 $\mu$m using the $hires$ method 
\citep{Menshchikov2021}, a generalization of the differential resolution 
enhancement algorithm described by \citet{Palmeirim2013}.
 From the filamentary structures obtained by \citet{Kumar2021}, we present skeletons of the filaments in which polarization segments are detected in Figure \ref{p2:fig:pol}. Skeletons of the filaments are obtained using the medial axis algorithm, which traces the locus of the set of all circles which have more than one tangent point on the boundary of the filamentary structure.
The skeletons show radial and spiral structures toward the center of Mon R2, IRS 1, and are similar to filamentary structures identified in C$^{18}$O molecular line data \citep{Trevino2019}.

We obtained magnetic and other physical properties toward 9 filaments among the 16 which are identified in the center of Mon R2 (these filaments and their properties are listed in Table \ref{p2:table:fila}). These nine filaments have a length greater than 0.45 pc, and sufficient numbers of polarization segments in each filament to obtain polarization angle dispersions. We adopted a uniform filament width of 0.1 pc in Mon R2, the width used by \citet{Kumar2021} to estimate filament masses. The width of 0.1 pc in Mon R2 cannot be resolved in our observations because of the limitation of the resolution. However, A width of 0.1 pc has been suggested to be the typical width of filaments in nearby star-forming regions based on \textit{Herschel} observations (\citealt{Arzoumanian2011}; \citealt{Palmeirim2013}; \citealt{Andre2014}; \citealt{Arzoumanian2019}) Moreover, \citet{Priestley2022} showed that magnetically sub-critical filaments have a universal width of 0.1 pc. All  of the filaments that we found in Mon R2 are found to be in a magnetically sub-critical state. For these reasons, we adopted a uniform width of 0.1 pc when calculating mean values of all parameters in each filament. We did not consider the inclination angle of Mon R2 when calculating these properties.

\begin{deluxetable*}{ccccccccc}
\tablecolumns{12}
\tablewidth{0pt}
\tablecaption{Physical parameters of filaments in Mon R2, and their uncertainties
\label{p2:table:fila}}
\tablehead{
(1) & (2) & (3) & (4) & (5) & (6) & (7) & (8) & (9)  \\
\colhead{} & \colhead{$M$} & \colhead{$L$} & \colhead{$M/L$} & \colhead{$<n(\mathrm{H}_2)>$} &\colhead{$<\Delta V>$} & \colhead{$<\sigma_\theta>$} &  \colhead{$<B_{\mathrm{POS}}>$}& \colhead{$<\lambda>$} \\
& \colhead{[$M_{\odot}$]} & \colhead{[pc]} & \colhead{[$M_\odot$ pc$^{-1}$]} & \colhead{ [10$^4$ cm$^{-3}$]} &\colhead{[km s$^{-1}$]} &  \colhead{[degree]}& \colhead{[mG]} & \colhead{}
}
\startdata
1  & 46 & 0.59 & 78 & 9 & 1.29$\pm$0.03 & 6.0$\pm$1.6 & 0.7$\pm$0.2 & 0.39$\pm$0.11\\
2  & 52 & 0.64 & 81 & 10 & 1.24$\pm$0.03 & 8.3$\pm$0.9 & 0.9$\pm$0.3 & 0.49$\pm$0.06\\
3  & 78 & 0.65 & 120 & 14 & 1.49$\pm$0.03 & 4.4$\pm$0.6 & 1.6$\pm$0.2 & 0.31$\pm$0.04\\
4  & 48 & 0.49 & 98 & 12 & 1.05$\pm$0.03 & 5.8$\pm$0.7 & 0.7$\pm$0.2 & 0.55$\pm$0.07\\
5  & 65 & 0.57 & 115 & 14 & 1.33$\pm$0.03 & 4.1$\pm$0.5 & 1.5$\pm$0.3 & 0.34$\pm$0.06\\
9  & 74 & 0.53 & 138 & 18 & 1.72$\pm$0.04 & 5.8$\pm$0.7 & 1.3$\pm$0.3 & 0.39$\pm$0.05\\
11  & 116 & 0.80 & 145 & 18 & 1.68$\pm$0.03 & 9.1$\pm$0.4 & 1.1$\pm$0.1 & 0.61$\pm$0.03\\
12  & 81 & 0.75 & 107 & 15 & 1.3$\pm$0.03 & 6.6$\pm$0.7 & 1.2$\pm$0.2 & 0.53$\pm$0.06\\
14  & 85 & 0.48 & 178 & 22 & 1.38$\pm$0.04 & 5.9$\pm$0.7 & 1.1$\pm$0.2 & 0.65$\pm$0.09\\
\enddata
\tablecomments{Columns: (1) Identification number of each filament. (2) H$_2$ mass of each filament in units of $M_\odot$. (3) Length of each filament in units of pc. (4) Mass per unit length of each filament in units of $M_\odot$ pc$^{-1}$. (5) Mean number density of each filament in units of 10$^{4}$ cm$^{-3}$. (6) Mean FWHM of the non-thermal component of the C$^{18}$O velocity dispersion of each filament in units of km s$^{-1}$. (7) Mean polarization angle dispersion of each filament in units of degrees. (8) Mean plane-of-sky (POS) magnetic field strength of each filament in units of mG. (9) Mean mass-to-flux ratio of each filament in units of a critical mass-to-flux ratio.}

\end{deluxetable*}

The mass of each filament ($M$) was estimated using the relation $\mu m_{\mathrm{H}}A\sum N_i(\mathrm{H}_2)$, where the mean molecular weight, $\mu$ is 2.8 (\citealt{Crutcher2004}; \citealt{Kauffmann2008}), 
$m_{\mathrm{H}}$ is the mass of a hydrogen atom, $A$ is the surface area of the filament, and $N_i$(H$_2$) is the column density of molecular hydrogen of each pixel. To compute the masses, we used the 18.2$''$ resolution column density image of the seperated component of filaments in Mon R2 \citep{Kumar2021} produced by $getsf$. The length of each filament, $L$, is defined as the length of its skeleton. The mass per unit length of a filament, $M/L$, is obtained by dividing its mass by the length of the filament.

\citet{Trevino2019} investigated the dynamic properties of filaments in Mon R2 using $^{13}$CO and C$^{18}$O (1-0) and (2-1) line data obtained using the IRAM 30m telescope. They smoothed their data to a velocity resolution $\sim$ 0.17 km\,s$^{-1}$, which is similar to that of our C$^{18}$O (3-2) line data obtained by HARP. Our analysis is focused on a $\sim$1 pc $\times$ 1 pc area around the central hub region of Mon R2, whereas \citet{Trevino2019} covered a $\sim$5 pc $\times$ 5 pc area. The main filaments obtained by \citet{Trevino2019} are up to 10 times longer than those obtained in this study. The masses per unit length ($M$/$L$) of the filaments in our study are larger than those found by \citet{Trevino2019}. \citet{Kumar2021} showed the number of filaments increases and $M$/$L$ decreases as the distance from the center of Mon R2, IRS 1, increases. They suggested that there is filament coalescence towards the center of the hub-filament structure. Due to this coalescence, filaments become shorter and their $M$/$L$ ratios become larger if they are located closer toward the center. The differences between these results may be because of the enhanced coalescence of filaments in the core hub region. Another reason for the difference is that most of the mass in filaments could be concentrated toward the center of the hub-filament structure due to the longitudinal flow of mass along the filaments. Therefore, the $M$/$L$ ratio could be larger in the central filaments than in the outer filaments. 

The large velocity gradient (LVG) model \citep{Tafalla1997} can be used to estimate volume density and kinetic temperature from spectral lines observed in multiple transitions. Several spectral lines have been used to estimate the volume densities in the region around the IRS 1 source at the center of Mon R2 \citep{Choi2000, Pilleri2013, Pilleri2014}. Because the size of the UC H{\sc ii} region around IRS 1 is $\sim$ 20$''$, we divided the volume density estimated at an offset of 20$''$ from IRS 1 by the column density interpolated at the same coordinates in order to estimate the depth of Mon R2. The depth estimated in this way is about 0.11 pc. We assumed a flattened hub structure for Mon R2, and so we divided all column densities by the depth to estimate volume densities. The $\Delta V$ value of a given pixel is the full width at half maximum (FWHM) of the non-thermal component of the C$^{18}$O (3-2) spectral line obtained by HARP on the JCMT, $\Delta V^2=\Delta V_{\mathrm{obs}}^2- 8\ln 2(kT_k/ m_{\mathrm{C}^{18}\mathrm{O}})$, where $\Delta V_{\mathrm{obs}}$ is the measured FWHM of the C$^{18}$O spectral line, $T_k$ is the kinetic temperature, and $m_{\mathrm{C}^{18}\mathrm{O}}$ is the mass of the C$^{18}$O molecule. The $T_k$ of Mon R2 has been measured using NH$_3$ (1,1) and (2,2) line data obtained using the Green Bank Observatory \citep{Keown2019}. \citet{Keown2019} generated a model of ammonia spectra under the assumptions of local thermodynamic equilibrium and a single velocity component along the line of sight, and used this model to fit their ammonia line data. We used the map of $T_k$ which they obtained to estimate the thermal component of the observed C$^{18}$O line. We estimated values of $\Delta V_{\mathrm{obs}}$ by fitting  each C$^{18}$O spectral line measurement with multiple Gaussian profiles. Most of the C$^{18}$O spectral line measurements are fitted with a single Gaussian profile. The rest, 5.8\% of the total number of measurements, are fitted with multiple Gaussian profiles. When a spectral line is fitted with multiple Gaussian profiles in a given pixel, we chose the component whose central velocity was the closest to that in the nearest pixel with a single Gaussian component within 30$''$ radius. If a pixel with a single Gaussian component does not exist within this radius, we exclude that pixel from the map and from further analysis in order to avoid ambiguity.
The right panel in Figure \ref{p2:fig:vol_vel} shows the map of velocity dispersion in Mon R2 obtained in this way. The estimated velocity dispersions from C$^{18}$O is comparable to those from NH$_3$ in Mon R2 \citep{Keown2019}
 
Polarization angle dispersions, $\sigma_\theta$, magnetic field strengths, $B$, and mass-to-flux ratios, $\lambda$ were estimated using the DCF method, and assuming a rotating axisymmetric magnetic field model. The detailed procedures for estimating $\sigma_\theta$ and the model are described in section \ref{p2:subsec:morphology}. The details of $B$ and $\lambda$ are given in section \ref{p2:subsec:mag} and \ref{p2:subsec:lambda}, respectively. We estimated their mean values in each of the filaments shown in Table \ref{p2:table:fila}. Uncertainties on the physical parameters listed in the table were determined by propagating errors; see \citet{Hwang2021} for a detailed description of estimation of uncertainties.

\section{Discussions}\label{p2:sec:dis}

\subsection{Mean magnetic field}\label{p2:subsec:morphology}

We assumed a large-scale mean magnetic field geometry in Mon R2 based on a model suggested by \citet{Wardle1990} in order to estimate magnetic field strengths using the DCF method. The DCF method is commonly used to estimate magnetic field strengths in star-forming regions. The method enables us to measure a magnetic field strength using three observed quantities: polarization angle dispersion, gas number density, and non-thermal velocity dispersion. This method assumes that the distortion of the field with respect to a large-scale mean magnetic field is caused by turbulence. The degree of the distortion can be estimated from polarization angle dispersion, assuming that polarized light comes from dust grains aligned with respect to the magnetic field \citep{Andersson2015, Lazarian2007}. However, the distortion of a magnetic field in a star-forming region can be caused not only by turbulence but also by other processes such as gravitational collapse, rotation, and outflows (e.g., \citealt{Beuther2020}). We used a rotating axisymmetric magnetic field model to approximate the field distortion by the other processes. This model can serve as a large-scale mean magnetic field in the context of the DCF method. We then subtracted the polarization angle of the model from the observed polarization angle in a given pixel to estimate the field distortion caused by turbulent motion only. In this section, we describe how we determine the large-scale mean magnetic field and polarization angle dispersion in Mon R2. 

The magnetic field morphology in Mon R2 looks like a spiral or pinwheel structure centered at the position of IRS 1, which could be caused by cloud-scale rotation on 1 pc scales. The hub of Mon R2 has a flattened sheet-like structure and its rotation is estimated from position-velocity (PV) diagrams of C$^{18}$O and $^{13}$CO ($J$ = 2-1) spectral line data \citep{Trevino2019}. The integrated C$^{18}$O and $^{13}$CO intensity maps show a ring-like structure in the hub, and the PV diagram along the structure shows a sinusoidal curve reminiscent of a rotational motion. 
These observational results motivate us to apply a rotating axisymmetric magnetic field model for the large-scale mean magnetic field to our observed magnetic field morphology in Mon R2.

In high-mass star-forming regions, there are a few observations which show spiral magnetic field structures. \citet{Beuther2020} observed magnetic fields in the G327.3 star-forming region, finding that the magnetic field morphology shows spiral and radial structures toward the center of the region, suggesting that the morphology is mainly controlled by gravity. \citet{Wang2020} showed a magnetic field morphology along a spiral-like structure in G33.92+0.11. The magnetic fields and the local gravity are aligned along the spiral arm, similarly to the filamentary structures seen in high-density regions. Magnetic field lines in the high-mass star-forming region IRAS 18089-1732, detected by ALMA, show a spiral rotating structure, dragging material toward a hot molecular core located at the center of the region \citep{Sanhueza2021}. These previously observed morphologies are similar to that of Mon R2.
 
We used the magnetic field model suggested by \citet{Wardle1990} (hereafter the WK model) to interpret the large-scale mean magnetic field in Mon R2. These authors made a rotating axisymmetric magnetic field model to estimate the magnetic field orientation in the circumnuclear disk (CND) of the Galactic center. The CND extends up to 10 pc in infrared and molecular line observations \citep{Genzel1987, Genzel1989}. The physical size of Mon R2 is a few pc, which is smaller than the CND. \citet{Wardle1990} assumed that magnetic field lines are initially axially symmetric and perpendicular to the disk. From previous and present polarization observations in Mon R2, we can assume magnetic field structures similar to those assumed by \citet{Wardle1990}. Magnetic field lines estimated by {\it Planck} are well-ordered and perpendicular to most of the filaments in Mon R2  \citep{Kumar2021}. These magnetic field lines can be interpreted as field lines which are perpendicular to the flattened structure of Mon R2. The magnetic field orientations which we obtained show a radially symmetric structure toward the IRS 1 source. For these reasons, we can apply the magnetic field line assumptions of \citet{Wardle1990}. 

The WK model produces magnetic field morphologies by changing inclination angles and other parameters. These parameters are azimuthal to vertical magnetic field ratio, radial to vertical magnetic field ratio, the thickness of the disk, and inflow to azimuthal velocity ratio. \citet{Wardle1990} found the best-fit model of the magnetic field orientations estimated from polarization observations of the CND of the Galactic center at 100 $\mu$m. \citet{Hsieh2018} compared the WK model with polarization data obtained by SCUPOL on the JCMT in the CND at 850$\mu m$. The observed results are found to be consistent with a model in which the azimuthal and radial components of the magnetic field are more dominant than the vertical component (gc1 model in \citealt{Wardle1990}).
Although the scale of Mon R2 is smaller than that of the CND, the WK model still seems appropriate to describe the mean magnetic field of Mon R2 because Mon R2 has a sheet-like flattened structure, rotating gas motion, and a spiral pattern in its magnetic field morphology.

\begin{figure*}[htb!]
\epsscale{1.1}
\plotone{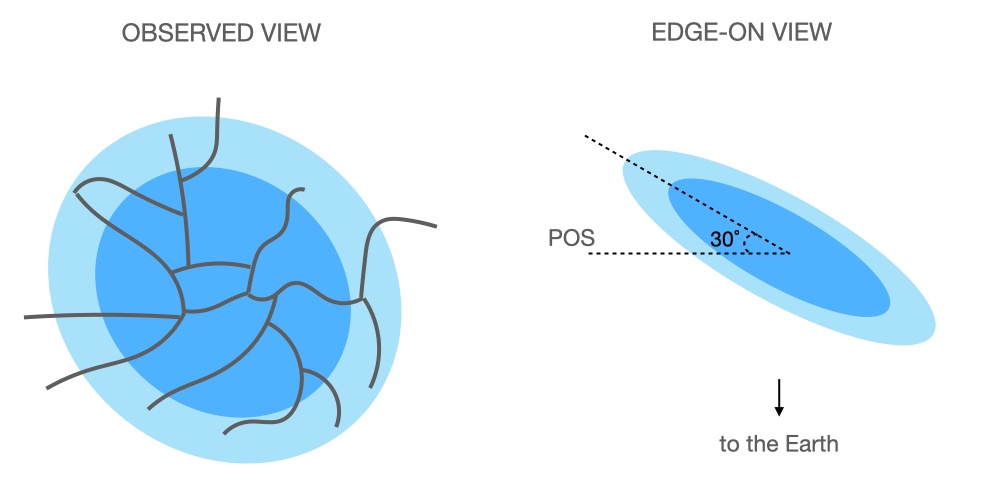}
\caption{A sketch of the observer's (left) and edge-on views of Mon R2  (right), with an inclination angle of 30$^{\circ}$. Black lines in the left panels show the filaments found in our observations. Black thick arrow in the right panel shows a direction towards an observer. Shapes in dark and light blue show the inner and outer parts of the hub, respectively.  
\label{p2:fig:view}} 
\end{figure*}

We found the best-fit model of the JCMT observations by changing the two parameters that determine the Stokes $Q$ and $U$ values in the WK model ($\omega$ and $\theta$, see Appendix \ref{p2:sec:wk}). $\omega$ is the angle between the direction of the three-dimensional (3D) magnetic field and the vertical axis from the plane of disk. The disk plane is perpendicular to the line of sight.
$\theta$ is the angle between the direction of a projected magnetic field line on the plane of the disk and the radial direction from the center of the disk to the line (for a schematic view, see Figure 4b in \citealt{Wardle1990}).
The inclination angle, $i$, is one of the important parameters for predicting magnetic field orientations from the model. We used the inclination angle of 30 degrees estimated by \citet{Trevino2019} in the WK model. In the derivation of inclination angle of the filaments, all the filaments are assumed to have the same velocity gradient in the matter flowing toward the hub. However, the observed velocity gradients in the filaments are found to be different, and this could be due to their different inclination angles. In this way the inclination angles of the three filaments going through the hub were estimated to have a mean value of 30 degrees. Based on this idea, \citet{Trevino2019} estimated inclination angles of the filaments.
Using $\omega$, $\theta$ and $i$, polarization segments projected on the plane of the sky (POS) are determined. Magnetic field orientations are obtained from the polarization segments by rotating them by 90 degrees.

We made a grid of magnetic field geometries using the WK model by changing $\omega$ and $\theta$, with the inclination angle $i$ fixed at 30 degrees. We calculated $\chi^2 = \sum{(\theta_{obs}-\theta_m)^2}/N$ for each pair of $\omega$ and $\theta$ values, where $\theta_{obs}$ and $\theta_m$ are polarization angles observed by the JCMT and predicted by the WK model, respectively, and $N$ is the number of measured angles. The estimated $\chi^{2}$ is at a minimum when $\omega$ and $\theta$ are 90 and -32 degrees, respectively. We thus used the WK model with these best-fitting parameters as the large-scale magnetic field in Mon R2.  
Additionally, we tested the dependence of the polarization angles predicted by the WK model on $i$, for the best-fitting values of $\omega$ and $\theta$ (Appendix \ref{p2:sec:wk}). We found that the polarization angles are not sensitive to $i$ if $i < 60$ degrees, and so our adopted inclination angle value of 30 degrees is reliable for our study.

\begin{figure*}[htb!]
\epsscale{1.0}
\plotone{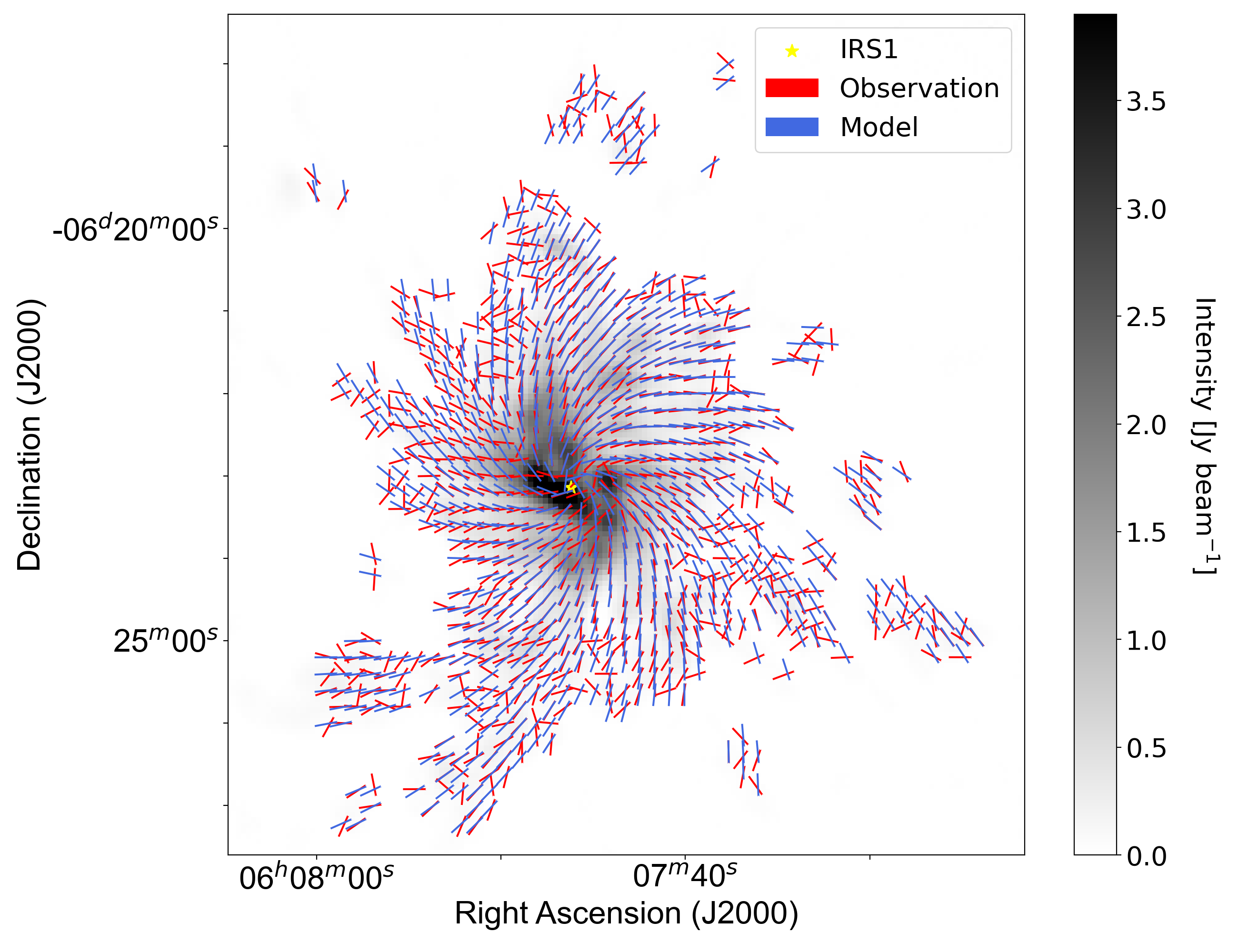}
\caption{Magnetic field orientations obtained from our polarization observations and from the best-fit WK model are shown as red and blue segments, respectively. The background gray-scale image shows the total intensity at 850 $\mu$m. The intensity scale of the background image is shown in the gray bar on the right side. A yellow star indicates the position of IRS 1 source. 
\label{p2:fig:mag_ori}} 
\end{figure*}

\begin{figure*}[htb!]
\epsscale{0.8}
\plotone{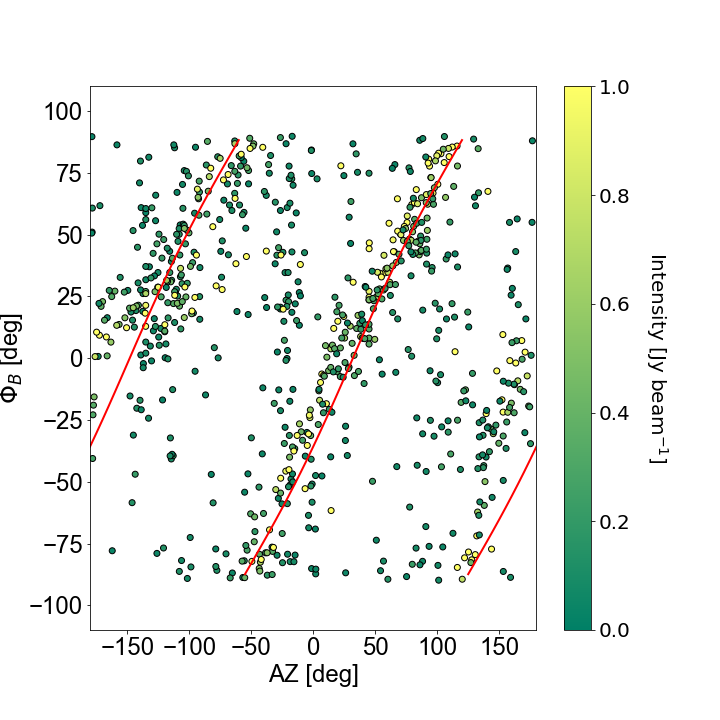}
\caption{Magnetic field angles, $\Phi_B$, as a function of azimuthal angle, AZ. The magnetic field and azimuthal angles are measured from West to North with respect to the origin at the position of IRS 1. The colored dots mark our polarization observations, and are color coded by total intensity. The color bar shows the mapping between color and total intensity. The red lines show magnetic field angles estimated using the best-fit WK model. When AZ is between 100 and 150 degrees, offsets between the observations and the best-fit WK model are apparent. \label{p2:fig:az_mag}} 
\end{figure*}

We chose to use the WK model based on our assumption that Mon R2 has a flattened structure, which was motivated by previous studies \citep{Trevino2019}. These authors presented a schematic view of Mon R2 consisting of a flattened hub with a radius of 1 pc and filaments extending up to 5 pc, based on their spectral line observations.  We observed the central 1 pc $\times$ 1 pc region, which is within the hub shown by \citet{Trevino2019}. This justifies our assumption that our observed region can be described by a flattened structure with an inclination angle of 30 degrees. Figure \ref{p2:fig:view} shows sketches of the observed view and the hypothesized edge-on view of the Mon R2 hub structure.
In the WK model, we focus not on the physical quantities, but rather on the magnetic field morphology predicted by the model. We used the axially-symmetric magnetic field structure predicted by the model to estimate the mean magnetic field geometry underlying our observations. The central part of Mon R2  (dark blue in Figure~\ref{p2:fig:view}) shows  good agreement between our observations and the model. However, the outer parts (light blue) show larger differences between the model and our observations. This might be caused by the outer filaments not being in the plane of the flattened structure. We excluded those outer parts which show large differences between the model and our observations when we  estimated magnetic field strengths, mass-to-flux ratios and Alfv\'en Mach numbers,  as discussed in Sections \ref{p2:subsec:mag}, \ref{p2:subsec:lambda} and \ref{p2:subsec:alfven}.

Figure \ref{p2:fig:mag_ori} shows the map of magnetic field orientations obtained from our POL-2 observations (red) and from the WK model (blue) with the best-fit parameters, $\omega=90^{\circ}$, $\theta=-32^{\circ}$ and $i=30^{\circ}$ degrees (hereafter the best-fit WK model) in Mon R2. The red and blue segments show spiral structures and are fairly well-matched to each other. The magnetic field orientations observed in the outskirts of Mon R2 are not ordered or matched with the model, which could be caused by low signal-to-noise ratio, lying on other plane compared to the central part of Mon R2, or by turbulence being dominant over magnetic fields in this region. There are deviations around IRS 1 between our observations and the model. Mangetic field lines around IRS 1 can be affected by the feedback from the central star or gravitational collapse. 
Figure \ref{p2:fig:az_mag} shows observed magnetic field orientations, $\Phi_B$, as a function of the azimuthal angle, as measured from the central position of Mon R2, IRS 1. The magnetic field orientations and azimuthal angles were calculated from West to North. The red lines indicate the angles of magnetic field segments as a function of azimuthal angle predicted by the best-fit WK model. The overall magnetic field angles are well-matched by the best-fit WK model, but structural offsets between observed magnetic angles and the best-fit WK model are apparent over a range of azimuthal angles from 100 to 150 degrees.

\begin{figure*}[htb!]
\epsscale{1.1}
\plotone{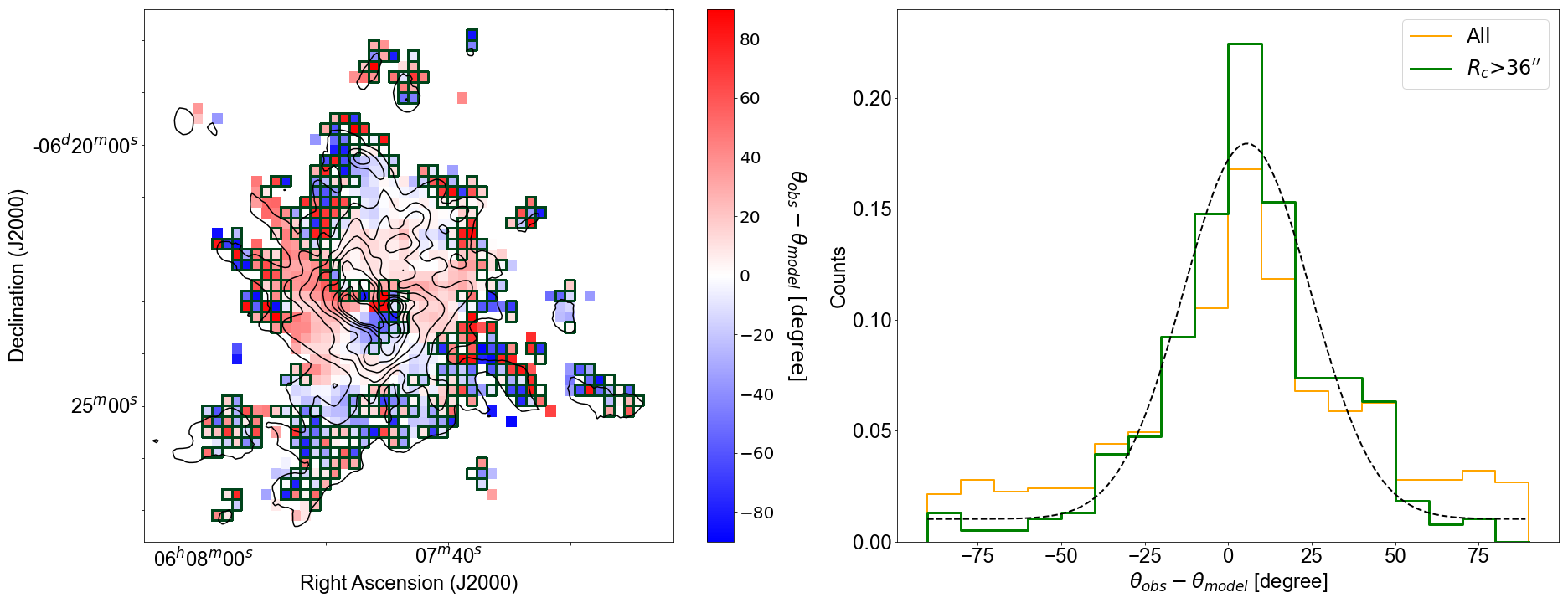}
\caption{Left panel: map of angle differences, $\Delta \theta$, between the POL-2 observations, $\theta_{obs}$, and the best-fit WK model, $\theta_{model}$. Positive and negative values mean that magnetic field orientations of POL-2 observations are directed counter-clockwise and clockwise from those of the best-fit WK model, respectively. The pixels outlined in green are excluded from our analysis due to high uncertainty, as their radii of curvature are smaller than the size of the box used in our analysis, 36$'' \times$ 36$''$ (see the Appendix \ref{p2:sec:curv} for details). Right panel: histograms of the angle differences. The orange histogram includes all pixels in the left panel. The green histogram excludes the pixels outlined with green in the left panel. The dashed line shows the Gaussian fitted to the green histogram, the standard deviation of which is 19 degrees. \label{p2:fig:obsdiff}} 
\end{figure*}

We constructed a map of the angle differences ($\Delta \theta$ = $\theta_{obs} - \theta_{model}$) between our two maps of the polarization angle: the WK model ($\theta_{model}$) and the observations ($\theta_{obs}$). Figure \ref{p2:fig:obsdiff} shows a map of and histograms of the angle differences ($\Delta \theta$) in the left and right panels, respectively. Positive and negative values define the relative directions of $\theta_{obs}$, as counter-clockwise and clockwise compared to $\theta_{model}$, respectively. The pixels outlined with green in the left panel are those in which the radii of curvature are smaller than our analyzed box size. The details are explained in the next paragraph and in Appendix \ref{p2:sec:curv}. The orange line in the right panel is the histogram of angle differences in all pixels of the left panel. The green histogram in the right panel is obtained by excluding the pixels outlined with green in the left panel. The dashed line is a Gaussian function fitted to the green histogram, with a standard deviation of 19 degrees. The overall angle differences excluding the pixels outlined in green show a Gaussian distribution. The best-fit WK model produces a reliable large-scale magnetic field in Mon R2. However, the angle differences in the eastern filament (Filament 1 in Figure \ref{p2:fig:pol}), center and outskirts of Mon R2 are still larger than 25 degrees. The mean fields in these regions are not fully subtracted, so we subtracted the remaining local mean fields using a method suggested by \citet{Hwang2021}.

\citet{Hwang2021} estimated maps of polarization angle dispersion using a small box. Their method is based on the assumption that if the box size is smaller than the radii of curvature of the magnetic field lines, the mean magnetic field direction will be uniform in the box. A radius of curvature is estimated by drawing a circle such that two polarization segments become tangent lines to the circle. They subtracted the mean field orientation from the observed field orientations in the box. By moving the box and repeating these processes, they could obtain a map of the angle differences between the mean field and the observed orientations. Then, they estimated polarization angle dispersion as the root-mean-squared (rms) value of the angle differences in the box centered at each pixel. We applied this method to Mon R2 in order to estimate angle dispersions.

\begin{figure*}[htb!]
\epsscale{1.0}
\plotone{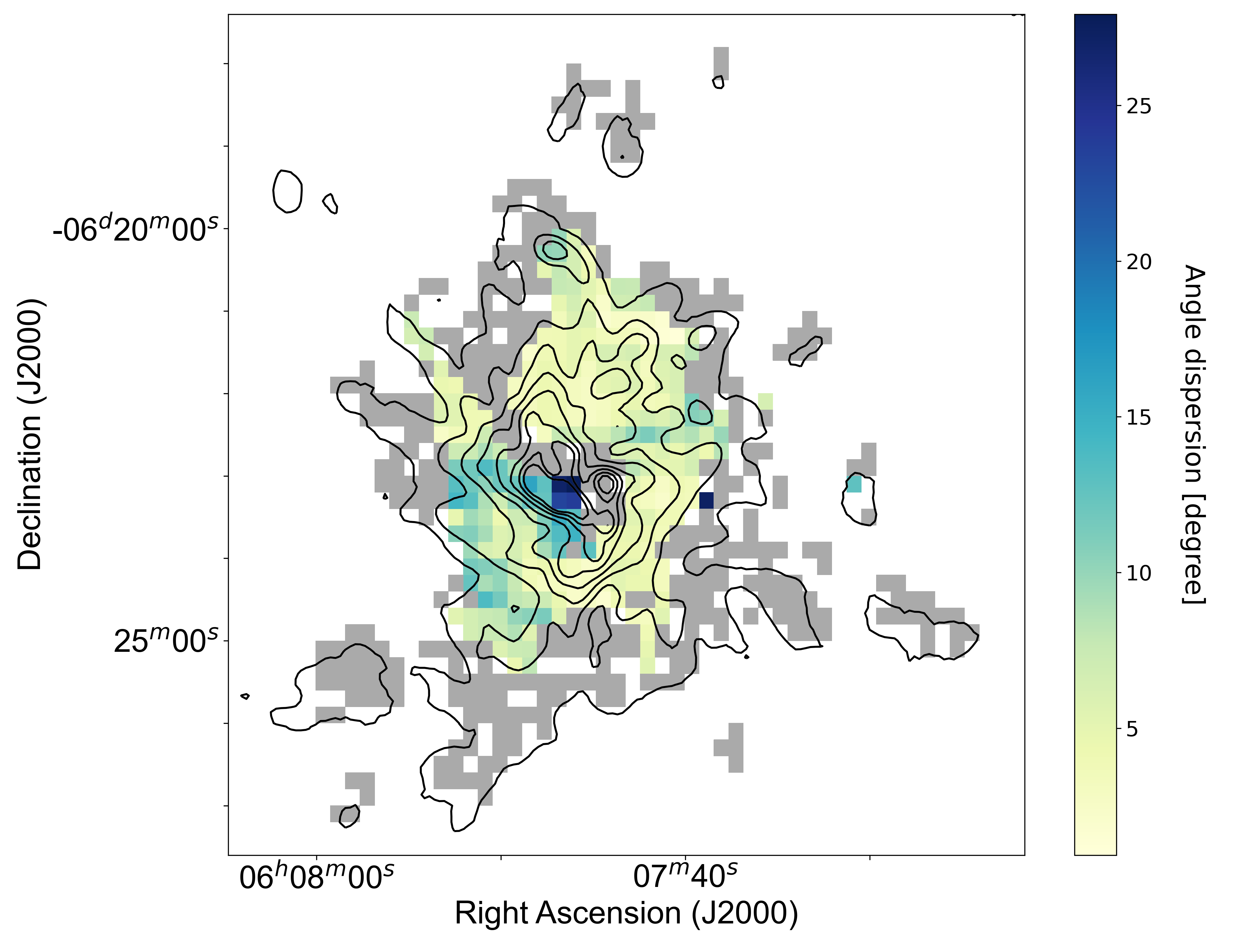}
\caption{A map of the angle dispersions of polarization segments. The WK model with $i=30^{\circ}$ was used to infer a large-scale mean magnetic field structure in Mon R2. A polarization angle dispersion in a gray pixel is highly uncertain, as a radius of curvature within the pixel is smaller than the size of box (see Appendix \ref{p2:sec:curv}). Contours are the same as those defined in Figure \ref{p2:fig:mom0}.
\label{p2:fig:ang_disp}} 
\end{figure*}

We estimated the dispersion of polarization angles by assuming that the WK model provides the large-scale mean magnetic field structure in Mon R2 (Figure \ref{p2:fig:ang_disp}). We took a small box, 36$'' \times$ 36$''$ (3 $\times$ 3 pixels), centered on a given pixel, and calculated a standard deviation of angle differences, ($\sum_{i=1}^M (\Delta \theta_i-<\Delta \theta>)^2/M)^{1/2}$, where $M$ and $<\Delta \theta>$ are  the number of segments and the mean value of the angle differences in the box. The standard deviation of the angle differences is taken to be the polarization angle dispersion in the center pixel of the box. By moving the small box and repeating this estimation of angle dispersions over Mon R2, we made a map of polarization angle dispersion (Figure \ref{p2:fig:ang_disp}). As shown in Figure \ref{p2:fig:ang_disp}, polarization angle dispersions in most regions of Mon R2 are smaller than 25 degrees, and so we were able to estimate magnetic field strengths using the DCF method \citep{Ostriker2001}. Mean angle dispersions in each filament are listed in Table \ref{p2:table:fila}. Each gray pixel shown in the figure has a radius of curvature that is smaller than the size of the box (Appendix \ref{p2:sec:curv}). If the box size, i.e., 3 $\times$ 3 pixels, is larger than the radius of curvature,  the angle dispersion of the polarization segments in the box will be overestimated \citep{Hwang2021}. We excluded these pixels when we estimated polarization angle dispersion in the box. Most gray pixels are located at the outskirt of Mon R2, which is due to an insufficient number of pixels and lower SNR in the outskirt regions. The central region of Mon R2 also shows small radii of curvature, which could result from the effect of the UC H{\sc ii} region, or from tangled magnetic fields in the region. UC H{\sc ii} regions can have magnetic field lines dragged along their boundaries which can make projected magnetic field lines appear to be changing direction rapidly (e.g., \citealt{Arthur2011}). We excluded the high-SNR pixels in the central regions which result from a tangled magnetic field in the line-of-sight (LOS) and POS. To estimate the angle dispersion in the outskirts and central region of Mon R2, we would need more observations with better sensitivity and higher angular resolution than the present observations have.
 
There are two differences in the method by which polarization angle dispersions are estimated between this work and \citet{Hwang2021}. \citet{Hwang2021} estimated  a mean field in each pixel, while we used the best-fit WK model. Additionally, \citet{Hwang2021} estimated polarization angle dispersion within the box as the rms value of angle differences. However, we estimated a standard deviation of angle differences in the box, because the angle differences within the small box can still contain a contribution from the mean field. If we did not use the best-fit WK model and instead applied the method of \citet{Hwang2021} to Mon R2, we could estimate angle dispersions in smaller regions of Mon R2 than we are able to when using the WK model. This is caused by the small radii of curvature of fields in Mon R2, and the obvious pattern of the magnetic field orientations in Mon R2.

\subsection{Magnetic field strengths}\label{p2:subsec:mag}

We used the DCF method to estimate magnetic field strengths. We obtained maps of column density, velocity dispersion, and angle dispersion as explained in sections \ref{p2:sec:res} and \ref{p2:subsec:morphology} (Figures \ref{p2:fig:vol_vel} and \ref{p2:fig:ang_disp}). We made the volume density map by dividing the column density map by a uniform depth, as discussed in section \ref{p2:sec:res}. We obtained interpolated maps of volume density and velocity dispersion, such that they are on the same pixel grid as the map of polarization angle dispersion. 
Then, we inserted the values of the three quantities in each pixel into the equation 
\begin{eqnarray} 
 B_{\text{pos}}\, [\mu \text{G}] = Q\sqrt{4\pi\rho}\frac{\sigma_{v}}{\sigma_{\theta}}   \nonumber \\
\approx 9.3\sqrt{n(\text{H}_2)\, [\mathrm{cm}^{-3}]}\frac{\Delta V\, [\mathrm{kms}^{-1}]}{\sigma_\theta \, [\mathrm{degree}]},
\label{p2:eq:cf}
\end{eqnarray}
where $B_{\text{pos}}$ is the magnetic field strength in the POS, $Q$ = 0.5 is the correction factor suggested by \citet{Ostriker2001} for the case where angle dispersion is less than 25 degrees, $\rho$ is the gas density, and $\sigma_{v}$ is the velocity dispersion. Other notation is explained in section \ref{p2:sec:res}. 
 
\begin{figure*}[htb!]
\epsscale{1.0}
\plotone{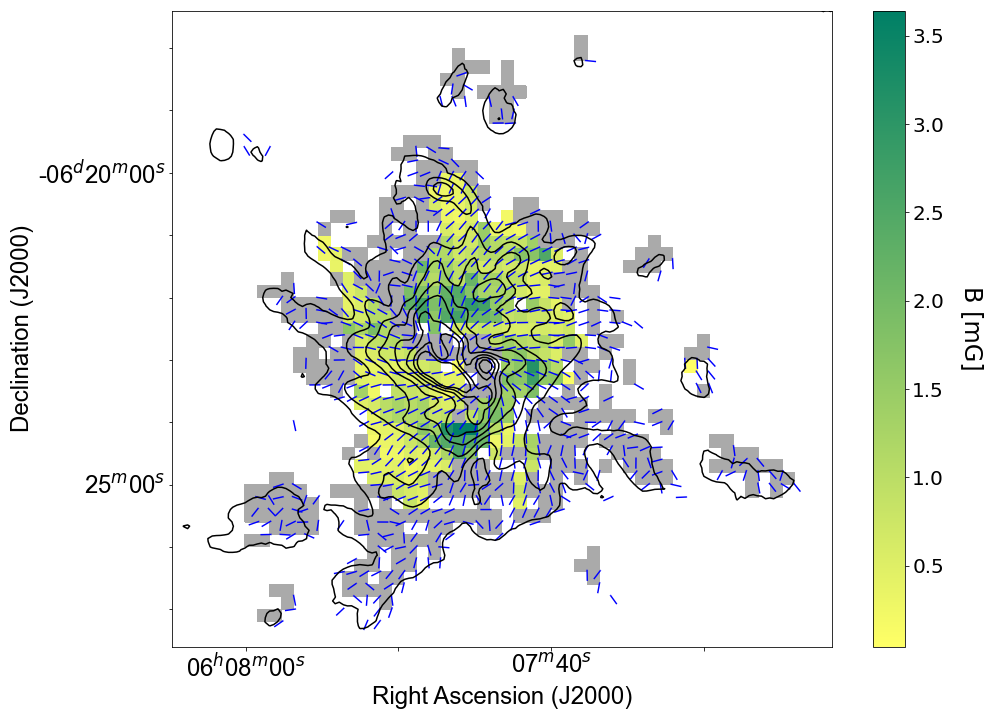}
\caption{A map of POS magnetic field strengths. The best-fit WK model is used as the large-scale mean magnetic field of Mon R2. Blue segments show the magnetic field orientations obtained from the POL-2 observations. Contours are the same as those defined in Figure \ref{p2:fig:mom0}. Gray pixels are excluded from our analysis as in Figure~\ref{p2:fig:ang_disp}. \label{p2:fig:mag_str}} 
\end{figure*}

Figure \ref{p2:fig:mag_str} shows the distribution of magnetic field strengths obtained using this implementation of the DCF method. The magnetic field strengths vary from 0.02 to 3.64 mG. The mean field strength is 1.0 $\pm$ 0.06 mG. Mean magnetic field strengths in each filament are listed in Table \ref{p2:table:fila}. The uncertainty on the magnetic field strength is estimated from the fractional uncertainties on velocity dispersion and polarization angle dispersion. The uncertainty on polarization angle dispersion is the  dominant term in the uncertainty on magnetic field strength. We estimated the uncertainty on polarization angle dispersion in the smoothing box to be the mean measurement uncertainty on polarization angle in that box. The uncertainty on velocity dispersion is taken to be the Gaussian fitting error of the C$^{18}$O lines. We did not include uncertainty on volume density, because we cannot determine the uncertainty on the depth. 
Magnetic field strengths in the outer regions are weaker than those in the center of Mon R2. This is mainly because the volume density is lower in the outer regions. Our result is the first measurement of the POS magnetic field strength distribution in the Mon R2 region using dust polarization observations. \citet{Knapp1976} determined LOS magnetic field strengths ranging from 0.003 to 0.4 mG using OH masers.
  
 \subsection{Mass-to-flux ratios}\label{p2:subsec:lambda}
 
The mass-to-flux ratio is used to estimate the relative importance of magnetic fields compared to gravity (\citealt{Mouschovias1976}; \citealt{Crutcher2004}). A dimensionless quantity, $\lambda$, is defined by dividing  the observed mass-to-flux ratio by its critical value. We used the critical value for a magnetized disk that is marginally supported by a magnetic field against gravity, ($M/\Phi)_{\mathrm{crit}}$ = 1/2$\pi G^{1/2}$, derived by \citet{Nakano1978}. \citet{Tomisaka2014} calculated a critical value for equilibria of isothermal filaments that have lateral magnetic fields. The two values are similar to each other, so we used the critical value estimated by \citet{Nakano1978}.

\begin{equation}
\lambda \equiv \frac{(M/\Phi)_{\text{obs}}}{(M/\Phi)_{\text{crit}}} = \frac{\mu m_{\text{H}} N(\text{H}_2)/B}{1/2\pi G^{1/2}} \nonumber \\
= 7.6 \times10^{-21} \frac{N(\text{H}_2)\, [\mathrm{cm}^{-2}] }{B\, [\mu \mathrm{G}]}, \label{p2:eq:mass_flux_ratio}                    
\end{equation}
 where $B$ is the strength of a 3D magnetic field and $G$ is the gravitational constant. When a molecular cloud is magnetically sub-critical ($\lambda$ $<$ 1), the magnetic field threading the cloud is strong enough to support the cloud against gravitational collapse. Conversely, a magnetically super-critical cloud ($\lambda$ $>$ 1) cannot resist gravitational collapse.  
 
\begin{figure*}[htb!]
\epsscale{1.0}
\plotone{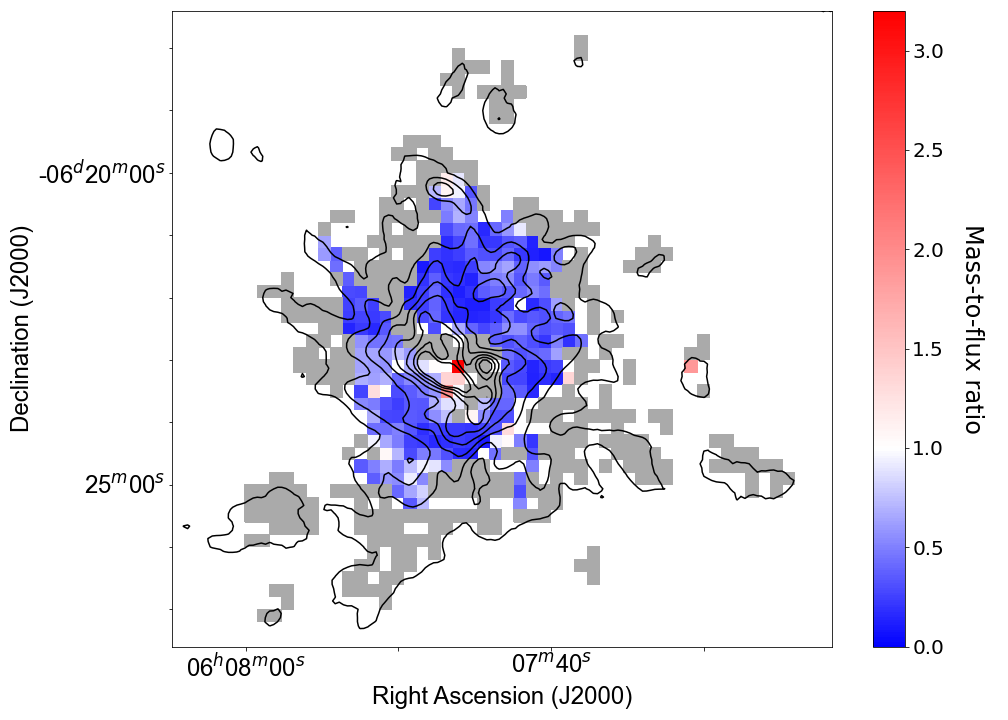}
\caption{A map of mass-to-flux ratios in Mon R2. The WK model with $i=30^{\circ}$ is used as the large-scale mean magnetic field of Mon R2. 
Contours and gray pixels are the same as in Figure~\ref{p2:fig:mom0} and \ref{p2:fig:ang_disp}, respectively. \label{p2:fig:mtob}} 
\end{figure*}

Figure \ref{p2:fig:mtob} shows the map of mass-to-flux ratios in Mon R2. The mass-to-flux ratios shown in the figure are estimated by substituting column densities and POS magnetic field strengths into equation (\ref{p2:eq:mass_flux_ratio}). The range of mass-to-flux ratios is from 0.09 to 3.21. Their mean and median values are 0.47$\pm$0.02 and 0.4. Most regions of Mon R2 are magnetically sub-critical, in which magnetic fields play an important role in supporting the cloud against gravitational collapse. However, we note that the uncertainties on magnetic field strengths obtained by the DCF method can be up to a factor of 4 \citep{Pattle2022} and we do not consider the 3D geometry of Mon R2. If we consider larger uncertainties on magnetic field strengths and mass-to-flux ratios, some regions can be magnetically trans-critical or super-critical. If we consider the 3D geometry of the flattened structure with an inclination angle of 30 degrees, we have to multiply the values in the figure by a factor of $\sin 30^{\circ} \cos 30^{\circ}$. Then, mass-to-flux ratios will be smaller than those in the figure, so most regions of Mon R2 are magnetically sub-critical. However, despite these uncertainties, the figure shows that magnetic fields in Mon R2 play an important role in resisting gravitational collapse.

The low star formation efficiency in Mon R2 could result from magnetic support against global collapse. \citet{Kumar2021} estimated a volume density profile from the center of Mon R2, the position of IRS 1, to which a power-law relation with an index of -2.17 is fitted in  the filamentary structures of Mon R2. This index is similar to that predicted from the condition for gravitational collapse, suggesting that Mon R2 hub may be in global collapse.
However, they estimated a star formation efficiency of a few \%, lower than the expected value of 20-30\% in Mon R2. They thus suggested that the low star formation efficiency in Mon R2 could be caused by magnetic fields supporting the cloud against global collapse. They showed  that the magnetic field orientations observed by {\it Planck} are perpendicular to and ordered along the Mon R2 cloud. In our results, the magnetic field orientations show a more complex structure, with higher resolution than those obtained by {\it Planck}. In spite of the complex morphology, most Mon R2 regions are magnetically sub-critical, meaning that magnetic fields are sufficiently strong to support the cloud against global collapse in these regions. Strong magnetic fields may affect the star formation efficiency by resisting gravitational collapse in Mon R2.

\subsection{Alfv\'en Mach number}\label{p2:subsec:alfven}

The Alfv\'en Mach number ($M_A$) is the ratio of turbulent velocity to Alfv\'en velocity, which is used to study the relative importance of turbulence and magnetic fields in molecular clouds \citep{Crutcher1999}. It is expressed as $M_A=\sqrt{3}\sigma_v/v_A$, where $v_A = B/\sqrt{4\pi\rho}$ is the Alfv\'en velocity. By using the measured inclination angle of 30 degrees for Mon R2, the 3D magnetic field strength is given by $B = B_{\mathrm{POS}}/\cos 30^{\circ}$. $M_A$ is thus given by $M_A=\sqrt{3}\sigma_\theta \cos 30^{\circ}/Q$, where $Q$ is the correction factor to the DCF method, 0.5. The sub-Alfv\'enic condition, $M_A <$ 1.0, indicates that magnetic pressure exceeds turbulent pressure.  Conversely, the super-Alfv\'enic condition, $M_A >$ 1.0, means that turbulent pressure exceeds magnetic pressure.

\begin{figure*}[htb!]
\epsscale{1.0}
\plotone{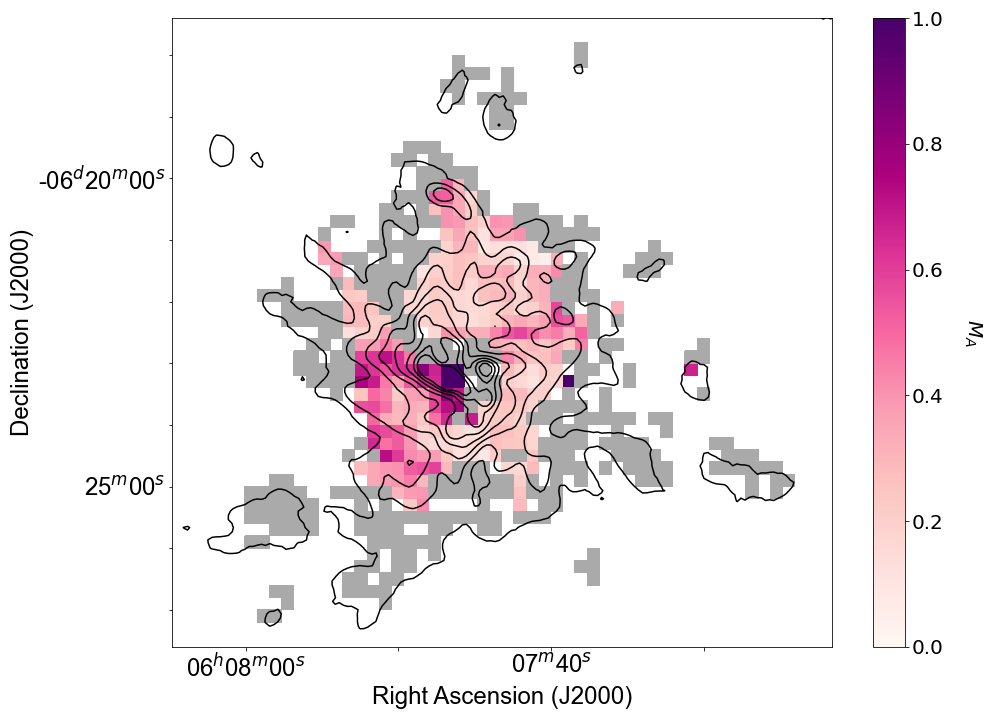}
\caption{A map of Alfv\'en Mach number in Mon R2. Contours and gray pixels are the same as in Figure~\ref{p2:fig:mom0} and \ref{fig:ang_disp}, respectively. \label{p2:fig:mach}} 
\end{figure*}
 
Figure \ref{p2:fig:mach} shows a map of the Alfv\'en Mach number in Mon R2. The range of $M_A$ is from 0.05 to 1.46, and its mean and median values are 0.35 $\pm$ 0.01 and 0.3, respectively. Most of the Mon R2 region is sub-Alfv\'enic, i.e. the magnetic pressure in the region is greater than the turbulent pressure. The magnetic field is thus relatively dominant compared to turbulence, and can regulate star-forming processes in most regions of Mon R2. The central region of Mon R2 is super-Alfv\'enic in which turbulence pressure is dominant compared to magnetic pressure.
 
\subsection{Magnetic and other physical properties of filaments}

\begin{figure*}[thb!]
\epsscale{1.0}
\plotone{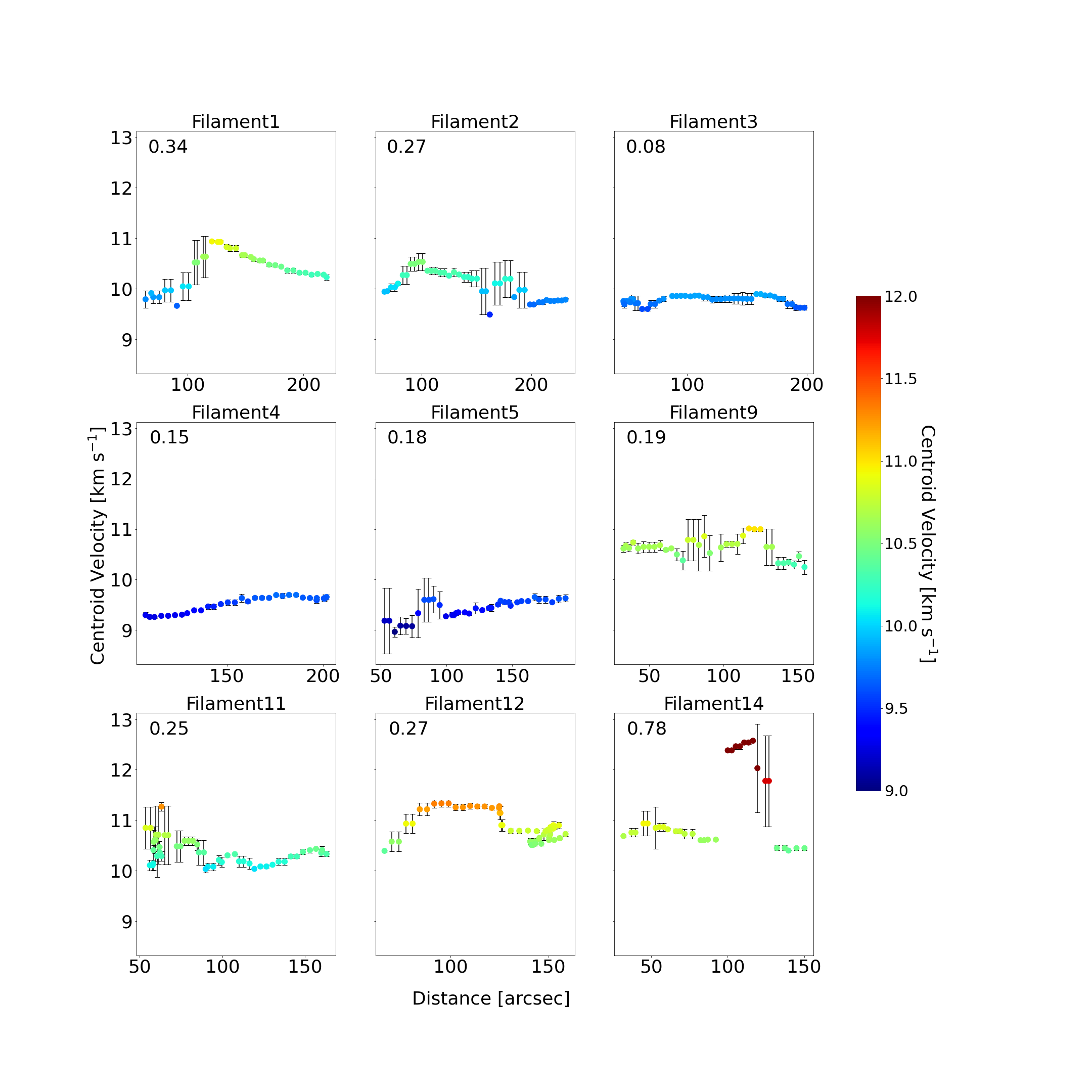}
\caption{The centroid velocities of the C$^{18}$O spectral line along each filament, which are interpolated at the coordinates of each skeleton. The starting distance of each filament is the filament pixel nearest to the IRS 1 source. The standard deviation of the centroid velocities of each filament in units of km\,s$^{-1}$ is provided in the upper left corner of each panel. \label{p2:fig:cent}} 
\end{figure*}

We chose the nine skeletons with lengths greater than 0.45 pc and analyzed them (Table \ref{p2:table:fila}). Figure \ref{p2:fig:cent} shows the centroid velocity of the C$^{18}$O spectral line as a function of distance from the IRS 1 source in each filament. The centroid velocities are taken to be the mean of the centroid velocities within the beam size of the JCMT at 850 $\mu$m, centered on each pixel along the length of the filaments. The error bar is the standard deviation of the centroid velocities within the beam size centered at each coordinate. The horizontal axes of each panel show angular distances between each pixel of the filaments and the IRS 1 source. We included the standard deviation of all centroid velocities in each filament in the upper left corner of each panel. Most of the filaments, except for filament 14, have standard deviations of less than 0.34 km s$^{-1}$. This value is about twice the spectral resolution of 0.15 km s$^{-1}$ and is significantly smaller than the typical line width of the C$^{18}$O spectral line, $\sim$1.5 km s$^{-1}$. We think that most of the filaments have coherent centroid velocity structure. There are a few points in filament 14 at which centroid velocities are larger than 11.5 km s$^{-1}$ (see Figure~\ref{p2:fig:cent}). This can be caused by overlapping two parts having different centroid velocities along a line-of-sight. Due to the points, the standard deviation of centroid velocities in the filament is 0.78 km s$^{-1}$.

\begin{figure*}[htb!]
\epsscale{1.0}
\plotone{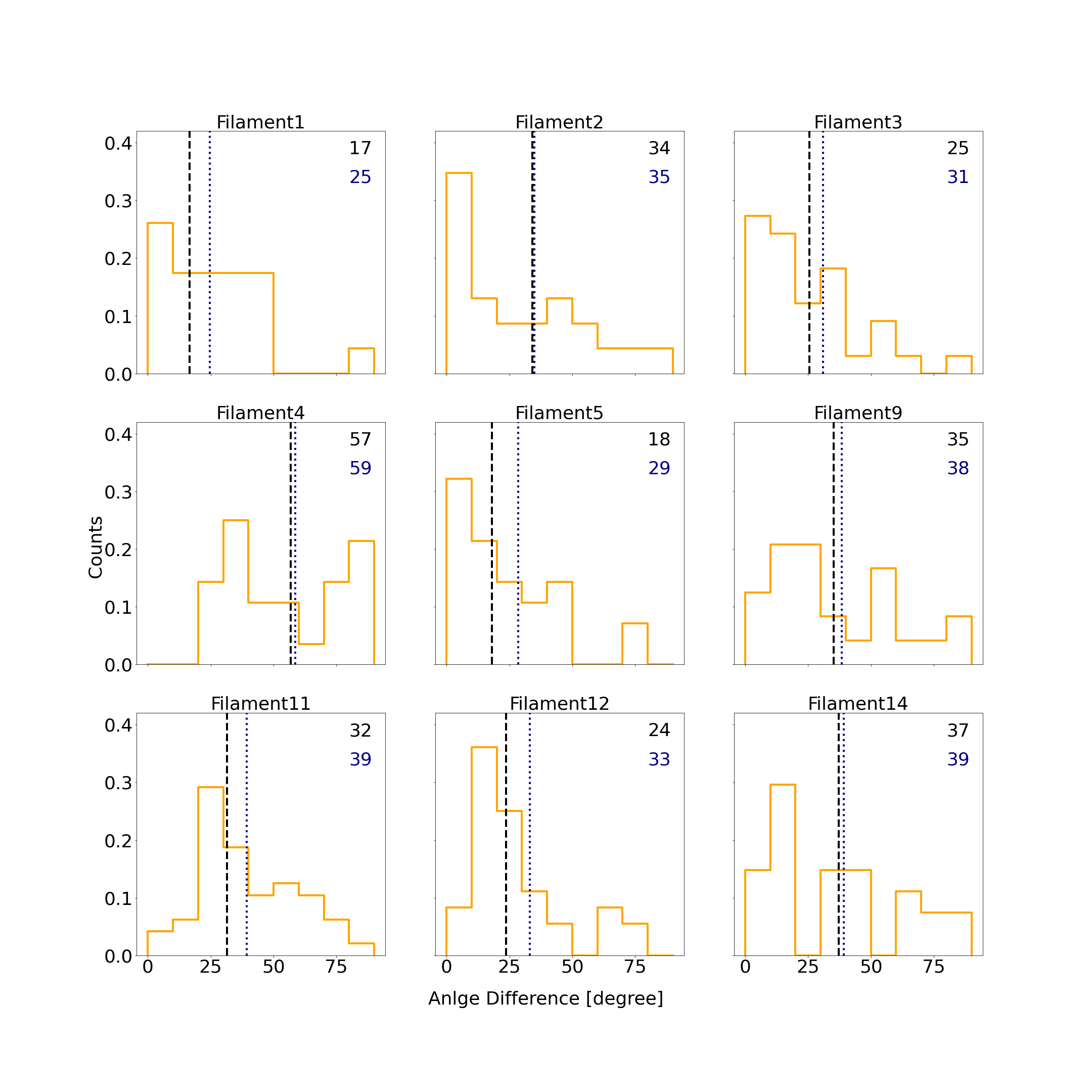}
\caption{Histograms of angle differences between filament angles and magnetic field orientations. The values written in black and navy colors in the upper right corner of each panel represent the mean and median values of the angle differences shown as the black dashed and navy dotted lines, respectively. \label{p2:fig:fila}} 
\end{figure*}

We compared the directions of the filament skeletons with magnetic field orientations. At each coordinate on a given skeleton, we estimated the angle between that coordinate and the adjacent one. The angle between the two coordinates is measured from North to East, the same convention as is used when measuring the polarization angle. Then, we calculated the angle difference between the filament angle and magnetic field orientation at each coordinate on the skeleton (Figure \ref{p2:fig:fila}). We calculated angle differences at all coordinates along each filament from one edge of the filament to the other. To clarify the relative orientations between filaments and magnetic fields, we estimated the median and mean values of the angle differences in each filament. These values are shown in the upper right corner of each panel, where black and navy colors represent median and mean values, respectively. In filament 4, the magnetic field orientations are preferentially perpendicular to the filament skeleton. However, other filaments are parallel to their local magnetic field orientations. Most filaments and magnetic fields show radial orientations, directed toward the center of the hub. This trend is consistent with the mass flows along filaments inferred using C$^{18}$O spectral line data by \citet{Trevino2019}. Magnetic fields can affect mass flows along filaments.

The magnetic field strengths and other physical parameters of nine filaments are listed in Table \ref{p2:table:fila}. We estimated all parameters in each pixel of Mon R2. We assumed the width of filaments to be 0.1 pc, and estimated mean values of each parameter within the filaments. Most of the filaments are magnetically sub-critical. Overall, in the Mon R2 region, magnetic fields are sufficiently strong to support filaments against gravitational collapse. The mass-to-flux ratios in the table are calculated using magnetic fields measured in the POS, and so could be changed by considering the 3D geometry and uncertainties mentioned in Section \ref{p2:subsec:lambda}. There is no correlation between these physical parameters and the angle differences of the filaments.
 
\section{Conclusions}\label{p2:sec:conc}
 
We have observed dust polarization and the C$^{18}$O $J=3-2$ spectral line in Mon R2 using SCUBA-2/POL-2 and HARP on the JCMT. The main results of our analysis of the role of magnetic fields in the hub-filament structure of Mon R2 are summarized as follows:

1. The distribution of polarization angles over Mon R2 shows a spiral magnetic field structure. These spiral magnetic fields converge on the center of Mon R2, in which the IRS 1 source is located. 

2. We estimated the magnetic and other physical properties  of 9 filaments obtained by \citet{Kumar2021}. The filaments are converging on the IRS 1 source. Their overall shape shows a spiral structure.

3. We compared the observations to a rotating axisymmetric magnetic field model. We showed that the overall magnetic field structure of Mon R2 is well represented by a magnetized rotating disk model.

4. We used this model to represent the mean magnetic field structure, and by subtracting it from the observed structure we estimated the angular dispersion of the magnetic field structure, which we used to calculate the  magnetic field strength. 

5. We obtained maps of angle dispersion, volume density and velocity dispersion in order to estimate magnetic field strengths in Mon R2 using the DCF method. After subtracting the mean field model, we calculated polarization angle dispersions using the method suggested by \citet{Hwang2021}. After obtaining volume density and velocity dispersion maps from dust continuum and C$^{18}$O observations, magnetic field strengths were derived in each pixel of the dust polarization map using the DCF method. The derived magnetic field strengths range from 0.02 to 3.64 mG, with a mean magnetic field strength of 1.0 $\pm$ 0.06 mG.

6. To dis`cuss the relative importance of gravity and magnetic fields, we estimated mass-to-flux ratios in Mon R2. The derived values of the mass-to-flux ratios in units of a critical value, $\lambda$, range from 0.09 to 3.21. The mean and median mass-to-flux ratios are 0.47$\pm$0.02 and 0.4, respectively. Most regions of Mon R2 are magnetically sub-critical.

7. We estimated Alfv\'en Mach numbers in Mon R2 in order to investigate the relative importance of turbulence and magnetic pressure. $M_A$ values range from 0.05 to 1.46, and their mean value is 0.35 $\pm$ 0.01, which means that magnetic pressure exceeds turbulence pressure in most of the Mon R2 region. The central region of Mon R2 is in a super-Alfv\'enic condition, in which turbulent pressure  dominates over magnetic pressure. 

8. We estimated centroid velocities along each filament. The filaments show velocity-coherent structures in which dispersions of centroid velocities are less than 0.34 km s$^{-1}$. We analyzed magnetic field orientations and strengths in nine filaments which converge on the center of Mon R2. Two filaments are perpendicular to their local magnetic field, while the other filaments are parallel to the local field orientations. All filaments are magnetically sub-critical. There are no correlations between magnetic and other physical properties.

\begin{acknowledgments}
The authors would like to thank the anonymous referee for his/her useful comments which helped to improve the draft.
We thank M. S. N. Kumar and A. Men'shchikov to provide maps of dust temperature, column density and filamentary structures for our analysis.
J.H. is supported by the UST Overseas Training Program 2022, funded by the University of Science and Technology, Korea (No. 2022-017).
K.P. is a Royal Society University Research Fellow, supported by grant number URF\textbackslash R1\textbackslash211322.
W.K. was supported by the National Research Foundation of Korea (NRF) grant funded by the Korea government (MSIT) (NRF-2021R1F1A1061794).
M.T. is supported by JSPS KAKENHI grant Nos. 18H05442, 15H02063, and 22000005. J.K. is supported by JSPS KAKENHI grant No. 19K14775. M.M. is supported by JSPS KAKENHI grant No. 20K03276. F.P. acknowledges support from the Spanish State Research Agency (AEI) under grant number PID2019-105552RB-C43.
C.E. acknowledges the financial support from grant RJF/2020/000071 as a part of a Ramanujan Fellowship awarded by Science and Engineering Research Board (SERB), Department of Science and Technology (DST), Govt. of India.
L.F. and F.K. acknowledge support from the Ministry of Science and Technology of Taiwan, under grant no. MoST107-2119-M-001-031- MY3 and from Academia Sinica under grant no. AS-IA-106-M03. L. F. acknowledges the support by the MoST in Taiwan through grants 111-2811-M-005-007 and 109-2112-M-005-003-MY3.
F.K. acknowledges support from the Spanish program Unidad de Excelencia Mar\'ia de Maeztu CEX2020-001058-M, financed by MCIN/AEI/10.13039/501100011033.
Y.S.D. would like to acknowledge the support from NSFC grants 12273051 and 10878003.
M.R. supported by the international Gemini Observatory, a program of NSF's NOIRLab, which is managed by the Association of Universities for Research in Astronomy (AURA) under a cooperative agreement with the National Science Foundation, on behalf of the Gemini partnership of Argentina, Brazil, Canada, Chile, the Republic of Korea, and the United States of America.
The JCMT is operated by the East Asian Observatory on behalf of The National Astronomical Observatory of Japan; Academia Sinica Institute of Astronomy and Astrophysics; the Korea Astronomy and Space Science Institute; the Operation, Maintenance and Upgrading Fund for Astronomical Telescopes and Facility Instruments, budgeted from the Ministry of Finance of China. Additional funding support is provided by the Science and Technology Facilities Council of the United Kingdom and participating universities and organizations in the United Kingdom, Canada and Ireland. Additional funds for the construction of SCUBA-2 were provided by the Canada Foundation for Innovation.
\end{acknowledgments}

\software{KAPPA \citep{Currie2008}, Starlink \citep{Jenness2013}}
\facilities{JCMT}

\appendix

\section{The dependence of predicted polarization angle on disk inclination}\label{p2:sec:wk}

\restartappendixnumbering 
 
\begin{figure*}[htb!]
\epsscale{0.6}
\plotone{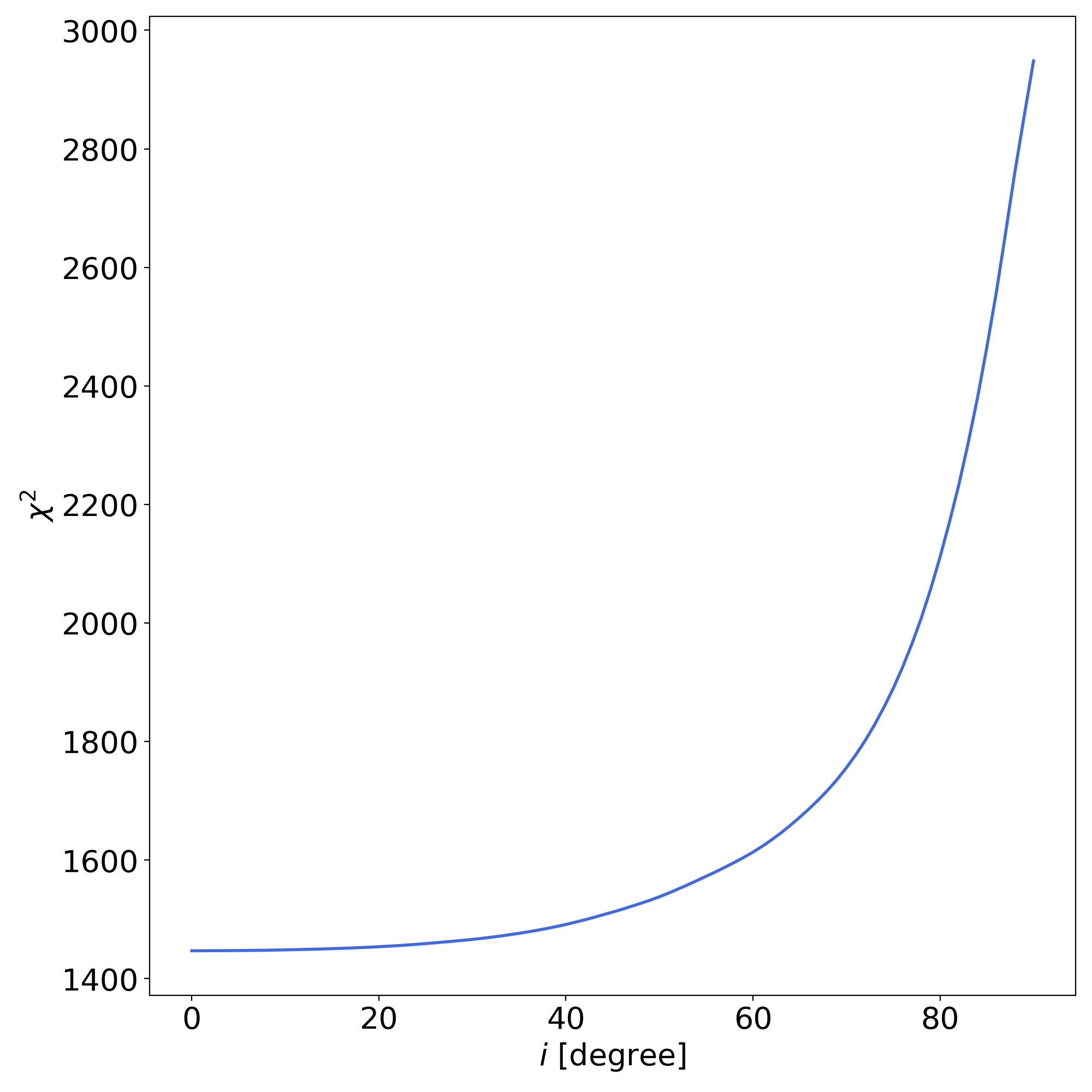}
\caption{The chi-squared values of the differences between the polarization angles observed by the JCMT and predicted by the WK model as a function of inclination angle. \label{p2:fig:inc}} 
\end{figure*}

\citet{Wardle1990} estimated magnetic field orientations in the CND of the Galactic center using an axisymmetric magnetic field model. They assumed axially symmetric magnetic fields in the disk. The polarization angle in the model ($\theta_m$) is calculated using the following equations \citep{Wardle1990}:

\begin{equation}
\cos 2\theta_m = \frac{q}{(q^2+u^2)^{1/2}}; \sin 2\theta_m = \frac{u}{(q^2+u^2)^{1/2}}, 
\label{p2:eq:cossin}
 \end{equation}
where their $q$ and $u$ are equivalent to our measured Stokes $Q$ and $U$ parameters. They are calculated using the following equations, 
\begin{equation}
q/N_d=\sin^2\omega[\cos^2(\theta+\phi)(\cos^2i+1)-1]+\cos^2\omega \sin^2i,
\label{p2:eq:q}
\end{equation}
\begin{equation}
u/N_d=\cos i \sin^2\omega \sin^2(\theta+\phi),
\label{p2:eq:u}
\end{equation}
where $N_d$ is dust column density, $\omega$ is the angle between the direction of the 3D magnetic field and the vertical axis extending from the plane of the disk, $\theta$ is the angle between the direction of a projected magnetic field line on the plane of the disk and the radial direction from the center of the disk to the line.
$\phi$ is the azimuthal angle from IRS 1, which is measured from West to North, and $i$ is the inclination angle. To calculate Stokes $Q$ and $U$, we used the same model parameters as the gc1 model, in which the vertical magnetic field component is less strong than the radial and azimuthal components \citep{Wardle1990}.  They used a fixed inclination value, $i=70^{\circ}$, for the CND, taken from the literature. 
We used $i=30^{\circ}$ for the Mon R2 estimated by \citep{Trevino2019} and obtained the best-fit values of $\omega=90^{\circ}$ and $\theta=-32^{\circ}$.  
We here checked the dependence of Stokes $Q$ and $U$ on the inclination angle. We calculated the polarization angles predicted by the WK model with $\omega=90^{\circ}$ and $\theta=-32^{\circ}$ by changing the inclination angle from 0 to 90 degrees in steps of 1 degree. 
Then, we calculated a chi-squared value, $\chi^2 = \sum{(\theta_{obs}-\theta_m)^2}/N$, where $\theta_{obs}$ and $\theta_m$ are the polarization angles observed by the JCMT and predicted by the WK model, and $N$ is the number of angles. Figure \ref{p2:fig:inc} shows $\chi^2$ as a function of inclination angle. When the inclination angle is less than 60 degrees, the resulting $\chi^2$ values are not sensitive to the inclination angle.

\section{Radii of curvature of polarization segments} \label{p2:sec:curv}

\begin{figure*}[htb!]
\epsscale{1.0}
\plotone{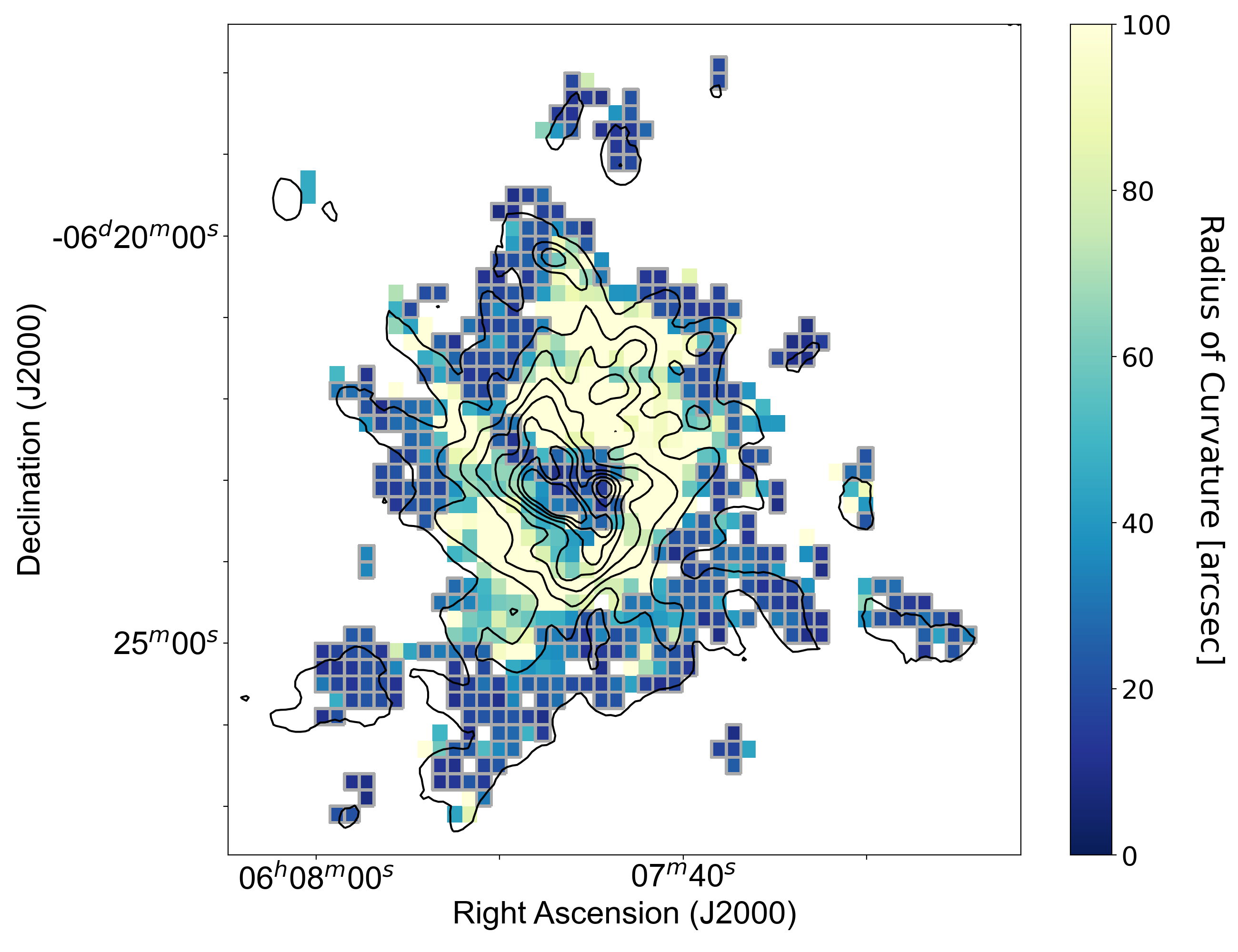}
\caption{A map of radii of curvature in units of arcseconds. A radius of curvature is obtained using the observed polarization segments at 850 $\mu$m, as explained in the main text. The gray outlined pixels have radii of curvatures smaller than 36 $''$. Most of these are located in the outskirts of the map. \label{p2:fig:curv}} 
\end{figure*}

We obtained the radius of curvature of the magnetic field in Mon R2 using polarization segments at 850 $\mu$m. To do so, we draw a circle going through two adjacent segments, which become tangent lines to the circle. The radius of the circle is also that of curvature (\citealt{Koch2012}; \citealt{Hwang2021}). For each pixel, we estimated radii of four circles with four pairs of segments, and obtained their mean value. The pairs are combinations of four segments located at $(i-k,j), (i+k,j), (i,j-k), (i,j+k)$ and that at $(i,j)$, where $i$, $j$ and $k$ are the right ascension and declination coordinates, and the angular size of a pixel, respectively.
We estimated polarization angle dispersion within a  3 $\times$ 3 pixel (36$'' \times$ 36$''$) box.
If the radius of curvature is smaller than the box size, it is known that the polarization angle dispersion in the box is overestimated \citep{Hwang2021}. A pixel that has a radius of curvature smaller than the box size should be excluded from the dispersion calculation. Those pixels are located in the outskirts and the center of the map as shown in Figure \ref{p2:fig:curv}.

\restartappendixnumbering 

\end{document}